\documentstyle[psfig]{mn}

\def\etal{et~al.}
\def\spose#1{\hbox to 0pt{#1\hss}}
\def\lta{\mathrel{\spose{\lower 3pt\hbox{$\mathchar"218$}}
     \raise 2.0pt\hbox{$\mathchar"13C$}}}
\def\gta{\mathrel{\spose{\lower 3pt\hbox{$\mathchar"218$}}
     \raise 2.0pt\hbox{$\mathchar"13E$}}}

\def\sextractor{{\sc sextractor}}

\def\dimsum{{\sc dimsum}}

\def\Ho50{$H_0 = 50$km\,s$^{-1}$\,Mpc$^{-1}$}

\def\ggal{{\rm gg}}
\def\rggal{{\rm rg}}
\def\cgal{{\rm cg}}
 
\title[Environments of 3CR radio galaxies]{The cluster environments of the
$\mathbf{z \sim 1}$ 3CR radio galaxies}

\author[P.~N.~Best]{P.~N.~Best\thanks{Present address: Institute for
Astronomy, Royal Observatory
Edinburgh, Blackford Hill, Edinburgh, EH9 3HJ}\thanks{Email:
pnb@roe.ac.uk}\\Sterrewacht Leiden, Postbus 9513, 2300 RA Leiden, the
Netherlands\\  
}
\pagerange{\pageref{firstpage}--\pageref{lastpage}}
\pubyear{1999}
\begin{document}
\label{firstpage}

\maketitle

\begin{abstract}
\noindent An analysis of the environments around a sample of 28 3CR radio
galaxies with redshifts $0.6 < z < 1.8$ is presented, based primarily upon
K--band images down to $K \sim 20$ taken using the UK Infrared Telescope
(UKIRT). A net overdensity of K--band galaxies is found in the fields of
the radio galaxies, with the mean excess counts being comparable to that
expected for clusters of Abell Class 0 richness. A sharp peak is found in
the angular cross--correlation amplitude centred on the radio galaxies,
which, for reasonable assumptions about the luminosity function of the
galaxies, corresponds to a spatial cross--correlation amplitude between
those determined for low redshift Abell Class 0 and Abell Class 1
clusters.

\noindent These data are complimented by J--band images also from UKIRT,
and by optical images from the Hubble Space Telescope. The fields of the
lower redshift ($z \lta 0.9$) radio galaxies in the sample generally show
well--defined near--infrared colour--magnitude relations with little
scatter, indicating a significant number of galaxies at the redshift of
the radio galaxy; the relations involving colours shortward of the
4000\AA\ break show considerably greater scatter, suggesting that many of
the cluster galaxies have low levels of recent or on--going star
formation. At higher redshifts the colour--magnitude sequences are less
prominent due to the increased field galaxy contribution at faint
magnitudes, but there is a statistical excess of galaxies with the very
red infrared colours ($J-K \gta 1.75$) expected of old cluster galaxies at
these redshifts.

\noindent Although these results are appropriate for the {\it mean} of all
of the radio galaxy fields, there exist large field--to--field variations
in the richness of the environments. Many, but certainly not all, powerful
$z \sim 1$ radio galaxies lie in (proto--)cluster environments.
\end{abstract}

\begin{keywords}
Galaxies: clustering --- Galaxies: active
\end{keywords}

\section{Introduction}
\label{intro}

Clusters of galaxies are the largest, most massive, collapsed structures
in the Universe, and as such are of fundamental importance for many
cosmological studies. They provide a unique probe of large--scale
structure in the early Universe and, as the systems which separated from
the Hubble flow at the earliest epochs, they contain the oldest
galaxies known. These can strongly constrain the first epoch of the
formation of ellipticals, and hence set a lower limit to the age of the
Universe; the effectiveness of using such old galaxies at high redshifts
towards this goal has been well demonstrated by Dunlop
et~al. \shortcite{dun96}. Further, because clusters contain large numbers
of galaxies at the same distance, they are important testbeds for models
of galaxy evolution.

The cores of optically--selected clusters are dominated by a population of
luminous early--type red galaxies which occupy a narrow locus in
colour--magnitude relations. Stanford, Eisenhardt and Dickinson
\shortcite{sta98} showed that, out to redshifts $z \sim 1$, the evolution
of the colours of early--type cluster galaxies on these relations is
completely consistent with passive evolution of an old stellar population
formed at high redshift. The small intrinsic scatter of the galaxy
colours, and particularly the fact that it remains small at redshifts $z
\gta 0.5$ \cite{sta98}, implies that the star formation of the ellipticals
comprising a cluster must have been well synchronized, and sets tight
limits on the amount of recent star formation that might have occurred
(e.g. Bower, Kodama and Terlevich 1998)\nocite{bow98}.

Cluster ellipticals also show a tight relationship between their effective
radius, effective surface brightness, and central velocity dispersion (the
`fundamental plane'; c.f Dressler \etal\ 1987, Djorgovski and Davies
1987).\nocite{dre87a,djo87b} The location of these ellipticals within the
fundamental plane has been shown to evolve with redshift out to $z =
0.83$, in a manner which implies that the mass--to--light ratio of the
galaxies evolves as $\Delta {\rm log} (M/L_{\rm r}) \sim -0.4 \Delta z$,
roughly in accordance with passive evolution predictions (e.g. van Dokkum
et~al 1998 and references therein).

These results are in qualitative agreement with `monolithic collapse'
models of galaxy formation, in which an elliptical galaxy forms the
majority of its stars in a single short burst of star formation at an
early cosmic epoch. On the other hand, the hierarchical galaxy formation
models favoured by cold dark matter cosmologies (e.g. White and Frenk
1991)\nocite{whi91} predict a later formation of galaxies with star
formation on--going to lower redshifts. These are supported by the
appearance of a population of bluer galaxies in many clusters at redshifts
$z \gta 0.3$ (the Butcher--Oemler effect; Butcher and Oemler
1978\nocite{but78}) and by the high fraction of merging galaxies seen in
the $z=0.83$ cluster MS1054$-$03 \cite{dok99a}. Since finding clusters at
high redshift is strongly biased towards the very richest environments, in
which early--type galaxies will have formed at the very highest redshifts,
hierarchical models can still explain the apparent passive evolution and
small scatter of the colour--magnitude relation out to redshifts $z \sim
1$.

It is clearly important to extend cluster studies out to still higher
redshifts, but the difficulty lies in the detection of clusters. At
optical wavelengths, the contrast of a cluster above the background counts
is minimal at these redshifts: the deep wide--area ESO Imaging Survey
(EIS) has found 12 `good' cluster candidates above redshift 0.8, but none
above $z=1.2$ \cite{sco99}. Addition of near--infrared wavebands helps
(e.g. Stanford \etal\ 1997)\nocite{sta97}, but is still a relatively
inefficient method. Selection using X--ray techniques is more reliable,
but X--ray surveys are currently sensitivity limited: the ROSAT Deep
Cluster Survey found about 30 clusters with $z \gta 0.5$, but none above
redshift 0.9 \cite{ros98}. Chandra and XMM will make a big improvement
here. An alternative approach is to use powerful radio galaxies as probes
of distant clusters: these can be easily observed out to the highest
redshifts, and there is growing evidence that at high redshifts they lie
in rich environments.

At low redshifts powerful double radio sources (FR\,II's; Fanaroff \&
Riley 1974)\nocite{fan74} are associated with giant elliptical galaxies
that are typically the dominant members of galaxy groups; the only nearby
radio source of comparable radio luminosity to the powerful high redshift
radio galaxies is Cygnus A, and this source lies in a rich cluster
\cite{owe97}.  At a redshift $z \sim 0.5$, analysis of the galaxy
cross--correlation function around FR\,II radio galaxies (Yates, Miller
and Peacock 1989)\nocite{yat89} and an Abell clustering classification
\cite{hil91} have shown that about 40\% of radio sources are located in
clusters of Abell richness class 0 or greater. At $z \sim 1$, the
circumstantial evidence that at least some powerful radio sources are
located at the centres of clusters is overwhelming and includes the
following:

\begin{itemize}
\item Detections of luminous X--ray emission from the fields of the radio
galaxies (e.g. Crawford and Fabian 1996)\nocite{cra96b}, sometimes
observed to be extended, indicating the presence of a relatively dense
intracluster medium.

\item Large over-densities of galaxies in the fields of some distant radio
sources, selected by infrared colour \cite{dic97a}.

\item Direct detections of companion galaxies with narrow--band imaging
(McCarthy, Spinrad and van Breugel 1995)\nocite{mcc95} and spectroscopic
studies \cite{dic97a}.

\item The observation that powerful radio galaxies are as luminous as
brightest cluster galaxies at $z \sim 0.8$, and have radial light profiles
which are well--matched by de Vaucouleurs law with large (10 to 15\,kpc)
characteristic radii (Best, Longair and R\"ottgering
1998b)\nocite{bes98d}.

\item The radio sources display large Faraday depolarisation and rotation
measures (e.g. Best \etal\ \shortcite{bes98a}; see also Carilli \etal\
\shortcite{car97} for redshift $z > 2$ radio sources), requiring a dense,
ionised surrounding medium.

\item Theoretical arguments that to produce such luminous radio sources
requires not only a high AGN power, due to a very massive central black
hole being fueled at close to the Eddington limit \cite{raw91b}, but also
a dense environment to confine the radio lobes and convert the jet kinetic
energy efficiently into radiation (e.g. see Barthel and Arnoud
1996).\nocite{bar96a}
\end{itemize}

At still higher redshifts, radio sources have been detected out to
redshift $z=5.2$ \cite{bre99}, and some well--studied sources are known to
lie in cluster environments. For example, towards the radio galaxy
1138$-$215 ($z=2.2$) extended X--ray emission has been detected (Carilli
\etal\ 1998) and narrow--band imaging reveals over 30 nearby Ly-$\alpha$
emitters \cite{kur00b}. Radio sources may therefore offer a unique
opportunity to study dense environments back to the earliest cosmic
epochs. As yet, however, there have been no {\it systematic} studies of
radio galaxy environments much beyond $z \sim 0.5$, and so it is important
to investigate in detail the nature of the environments of the general
population: do all powerful distant radio galaxies lie in cluster
environments, or do only a minority but these have `grabbed the
headlines'?; what are the properties of any clustering environments
(richness, radius, shape, etc.) surrounding these objects?; what is the
nature of the constituent galaxies (morphological composition,
segregation, etc.) of any detected cluster?

In this paper, deep near--infrared observations of the fields of a sample
of 28 powerful radio galaxies with $0.6 < z < 1.8$ are analysed, in
conjunction with Hubble Space Telescope (HST) images of the same fields,
to investigate the ubiquity and richness of the environments of the radio
galaxies. The observations, data reduction, and source extraction are
described in Section~\ref{obssect}. In Section~\ref{galcounts} the
integrated galaxy counts are considered, in Section~\ref{crosscor} the
angular and spatial cross--correlation amplitudes are derived, and in
Section~\ref{colmagsect} an investigation of the colour--magnitude and
colour--colour relations is carried out. The implications of the results
are discussed in Section~\ref{discuss}. Throughout the paper values for
the cosmological parameters of $\Omega = 1$ and $H_0 =
50$\,km\,s$^{-1}$Mpc$^{-1}$ are assumed.

\section{Observations and Data Reduction}
\label{obssect}

\subsection{The dataset}
\label{dataset}

The data used for this research was presented and described by Best,
Longair and R{\"o}ttgering \shortcite{bes97c}. In short, the sample
consists of 28 radio galaxies with redshifts $0.6 < z < 1.8$ drawn from
the revised 3CR sample of Laing, Riley and Longair \shortcite{lai83}. The
fields of these radio galaxies were observed at optical wavelengths using
the Wide-Field Planetary Camera II (WFPC2) on the HST generally for one
orbit in each of two different wavebands. They were also observed at
near--infrared wavelengths using IRCAM3 on the UK Infrared Telescope
(UKIRT) in the K--band for approximately 54 minutes, and in 20 of the 28
cases also in the J--band.

For 5 sources (3C13, 3C41, 3C49, 3C65, 3C340) a further 3 to 4 hours of
K--band observations have subsequently been taken using IRCAM3 in
September 1998 (Best \etal, in preparation), and these were combined with
the original data to provide much deeper images. These further data were
taken in the same manner as the original IRCAM3 data except for the use of
the tip-tilt system that was available on UKIRT then, but wasn't available
for the original runs; use of this system provided a significant reduction
in the effective seeing.

The HST data were reduced according to the standard Space Telescope
Science Institute (STScI) calibration pipeline \cite{lau89}, following the
description of Best \etal\ \shortcite{bes97c}.  The UKIRT data were
reduced using IRAF and mosaicked using the \dimsum\footnote{\dimsum\ is
the `Deep Infrared Mosaicking Software' package developed by Eisenhardt,
Dickinson, Stanford and Ward.} package, again following the general
procedure outlined by Best \etal\ \shortcite{bes97c}. During the
mosaicking process, the individual galaxy frames were block replicated by
a factor of 4 in each dimension to allow more accurate alignment. The
final images were then block--averaged in 2 by 2 pixels, resulting in
frames with a pixel size of 0.142 arcseconds. The seeing varied from image
to image, from about 0.7 to 1.2 arcsec. \dimsum\ also produces an output
exposure map indicating the exposure for each pixel on the combined image;
this is useful for weighting the different regions of the images, which
reach different limiting sensitivities owing to the dithering process used
in taking the data (see Best \etal\ 1997).\nocite{bes97c}

This dataset has a number of advantages and disadvantages for use in
studies of clustering around distant radio galaxies. On the negative side,
since these data were originally taken with the goal of studying the host
galaxies of the radio sources, no comparison fields were taken with which
the data can be compared to statistically remove background counts.
Further, IRCAM3 is only a 256 by 256 array with a field of view per frame
of just over 70 by 70 arcseconds.  After mosaicking this provided a final
image of nearly 100 by 100 arcsec, which corresponds to a little over
800\,kpc at a redshift of one for the adopted cosmology, but the highest
sensitivity is only obtained in the central 50 by 50 arcsec region. This
is a significantly smaller size than a cluster would be expected to have
at this redshift, and so only those galaxies in the central regions of any
prospective clusters will be investigated (cf. the much larger extent of
the structure around the radio galaxy 3C324 at redshift 1.206, as
demonstrated in Figure~2 of Dickinson \shortcite{dic97a}, although just
over half of the associated cluster members found there would lie within
the IRCAM3 field of view).

Although these two factors limit the usefulness of the dataset, they are
far outweighed by the benefits. First, the large size and essentially
complete nature of the sample make it ideal for surveying the average
environments of these sources. Second, the multi--colour nature of the
dataset provides important information on any cluster membership through
the creation of colour--magnitude diagrams and the determination of
photometric redshifts. Third, the availability of near--infrared data is
essential for such studies at these high redshifts since it continues to
sample the old stellar populations longwards of the 4000\AA\ break; in
the K--waveband the K and evolutionary corrections are small and
relatively independent of morphological type, even at redshifts $z \gta
1$, in contrast to the situation at optical wavelengths. Fourth, the
availability of the high resolution HST data allows accurate star--galaxy
separation down to the faintest magnitudes studied. For these reasons,
this dataset is well--suited to investigating the environments of $z \sim
1$ radio galaxies.

\subsection{K--band image detection and photometry}
\label{sextract}

Throughout these analyses, the K--band data were used as the primary
dataset. Image detection and photometry on the K--band frames was carried
out using \sextractor\ version 2.1 \cite{ber96}. As a result of the
dithering technique used to obtain near--infrared data, the exposure time
varies with position across the image meaning that the rms background
noise level also varies.  \sextractor\ allows the supply of an input
weight map by which the local detection threshold is adjusted as a
function of position across the image to compensate the varying noise
levels, thus avoiding missing objects in the most sensitive regions of the
image or detecting large numbers of spurious features in the noisiest
regions. The output exposure map of each field produced by \dimsum\ was
used as the weight map for \sextractor.  The source extraction parameters
were set such that, to be detected, an object must have a flux in excess
of 1.5 times the local background noise level over at least $N$ connected
pixels, where $N$ was varied a little according to the seeing conditions,
but in general was about 20 (equivalent to 5 pixels prior to the block
replication and averaging during the mosaicking procedure).

To test the validity of this extraction method, a search for negative
holes was carried out using the same extraction parameters, and resulted
in a total of only 43 negative detections throughout the 28 images, that
is, about 1.5 negative detections per frame. The fluxes associated with
these negative detections all correspond to positive features below the 50\%
completeness limit and so it is expected that there are essentially no
false positive detections above that limit.

The output from \sextractor\ was examined carefully, with minor (typically
1--2 per field) modifications being made to the catalogue if necessary:
objects coinciding with the spikes of bright stars were removed, the data
for occasional entries that had been separated into two objects by
\sextractor\ but were clearly (on comparison with data at other wavebands) a
single object were combined, and very occasionally it was necessary to add
to the catalogue an obvious object which had been missed due to its
proximity to a brighter object.  Further, all objects that lay within 21
pixels (3 arcsec) of the edge of an image were removed from the
catalogue in case their magnitudes were corrupted by aperture truncation.
\sextractor\ provides a {\sc flag} parameter to indicate the reliability of
its measured magnitudes. After this trimming of objects close to the edges
of the fields, less than 2\% of all objects had high {\sc flag} values
indicative of truncated apertures, saturated pixels, or corrupted
data. These objects were also removed from the catalogue.

\sextractor's {\sc mag\_best} estimator was used to determine the
magnitudes of the sources; this yields an estimate for the `total'
magnitude using Kron's \shortcite{kro80} first--moment algorithm, except
if there is a nearby companion which may bias the total magnitude estimate
by more than 10\% in which case a corrected isophotal magnitude is used
instead.  The determined magnitudes were also corrected for galactic
extinction, using the extinction maps of Burstein and Heiles
\shortcite{bur82a}.

To investigate the accuracy of these total magnitudes, and the
completeness level of the source extraction as a function of position on
the image, Monte--Carlo simulations were carried out using the following
4--step process.

\noindent {\it Step 1:} The point--spread function (PSF) of each K--band
image was determined using objects which were unsaturated and which were 
unresolved on the HST images.

\noindent {\it Step 2:} A series of model galaxies was made. 30\% of the
galaxies were assumed to be ellipticals, with radial light profiles
governed by de Vaucouleurs' law, $I(r) \propto {\rm exp} \left [-7.67
\left((a/a_{\rm e})^2 + (b/b_{\rm e})^2\right )^{1/8}\right ]$, where $a$
and $b$ are the distances along the projected major and minor axes, and
$a_{\rm e}$ and $b_{\rm e}$ are the characteristic scale lengths in those
directions. Lambas, Maddox and Loveday \shortcite{lam92} provide a
parameterization for the ellipticities of elliptical galaxies in the APM
Bright Galaxy Survey in terms of the parameter $p = b_{\rm e}/a_{\rm e}$.
An ellipticity was drawn from each galaxy at random from this
distribution, and the position angle of the major axis was also chosen
randomly. Songaila \etal\ \shortcite{son94} found that for galaxies with
$18 < K < 19$ the median redshift is about 0.6, which means that for $K
\gta 18$ (where detection and completeness are being tested) the majority
of the galaxies will be distant enough that their angular sizes are
relatively insensitive to redshift. High redshift ellipticals have typical
characteristic sizes ranging from 2 to 10\,kpc (e.g. Dickinson
1997)\nocite{dic97a}, and so the apparent characteristic size, $r_{\rm e}$
($= (a_{\rm e}b_{\rm e})^{1/2}$), of each elliptical was chosen randomly
from the range 0.2 to 1.2 arcsec.

The remaining 70\% of the galaxies were built using a single exponential
profile appropriate for galaxy disks, $I(r) \propto {\rm exp} [-r/r_{\rm
d}]$, where $r_{\rm d}$ is the characteristic scale length of the
disk. Mao, Mo and White \shortcite{mao98} show that the characteristic
scale lengths of disks decreases with redshift approximately as
$(1+z)^{-1}$, but Simard \etal\ \shortcite{sim99c} shows that a
magnitude--size correlation somewhat counter--balances this: at a given
apparent magnitude, higher redshift objects must be intrinsically brighter
and so are larger. From their data, the typical scale length of galaxy
disks at around our completeness limit will be 1--4\,kpc. The disk scale
length of each galaxy was therefore chosen at random from the range 0.1 to
0.5 arcsec. The disk inclination was chosen at random from 0 to 90
degrees, with inclinations greater than 75 degrees being replaced by 75
degrees to account for the non--zero thickness of the disk. The projected
orientation of the galaxy was also chosen randomly.

The model galaxies were then convolved with the stellar PSF derived in
step 1.

\noindent {\it Step 3:} Five stellar or galactic objects were added to each
frame with a random position and scaled to a random magnitude in the range
$15 < K < 22$. \sextractor\ was then run on the new image using the same
input parameters as for the original source extraction, to see if the
added objects were detected, and if so to determine the difference between
the total magnitude measured by \sextractor\ and the true total magnitude.

\noindent {\it Step 4:} Step 3 was repeated until 25000 stellar objects
and 25000 model galaxies had been added to each image.

\begin{figure}
\centerline{
\psfig{file=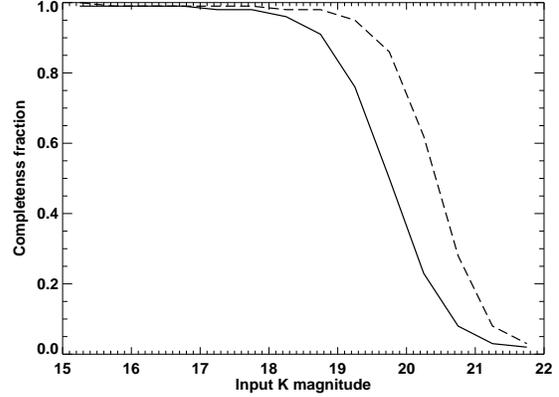,angle=90,width=7.8cm,clip=}
}
\caption{\label{compplot} The completeness fraction for both stellar
(dashed line) and galactic (solid line) model objects as a function of
input magnitude, averaged over all 28 fields.}
\end{figure}

From these results, the mean completeness fraction was determined as a
function of magnitude for both stars and galaxies over the entire set of
images. The results are shown in Figure~\ref{compplot}. The 50\%
completeness limits are $K=19.75$ for galaxies and $K=20.4$ for stars. As
discussed in Section~\ref{dataset}, new deeper observations are available
for five fields, and considering these deeper fields alone, the 50\%
completeness limits are 20.2 and 20.9 for galaxies and stars respectively.

\begin{figure}
\centerline{
\psfig{file=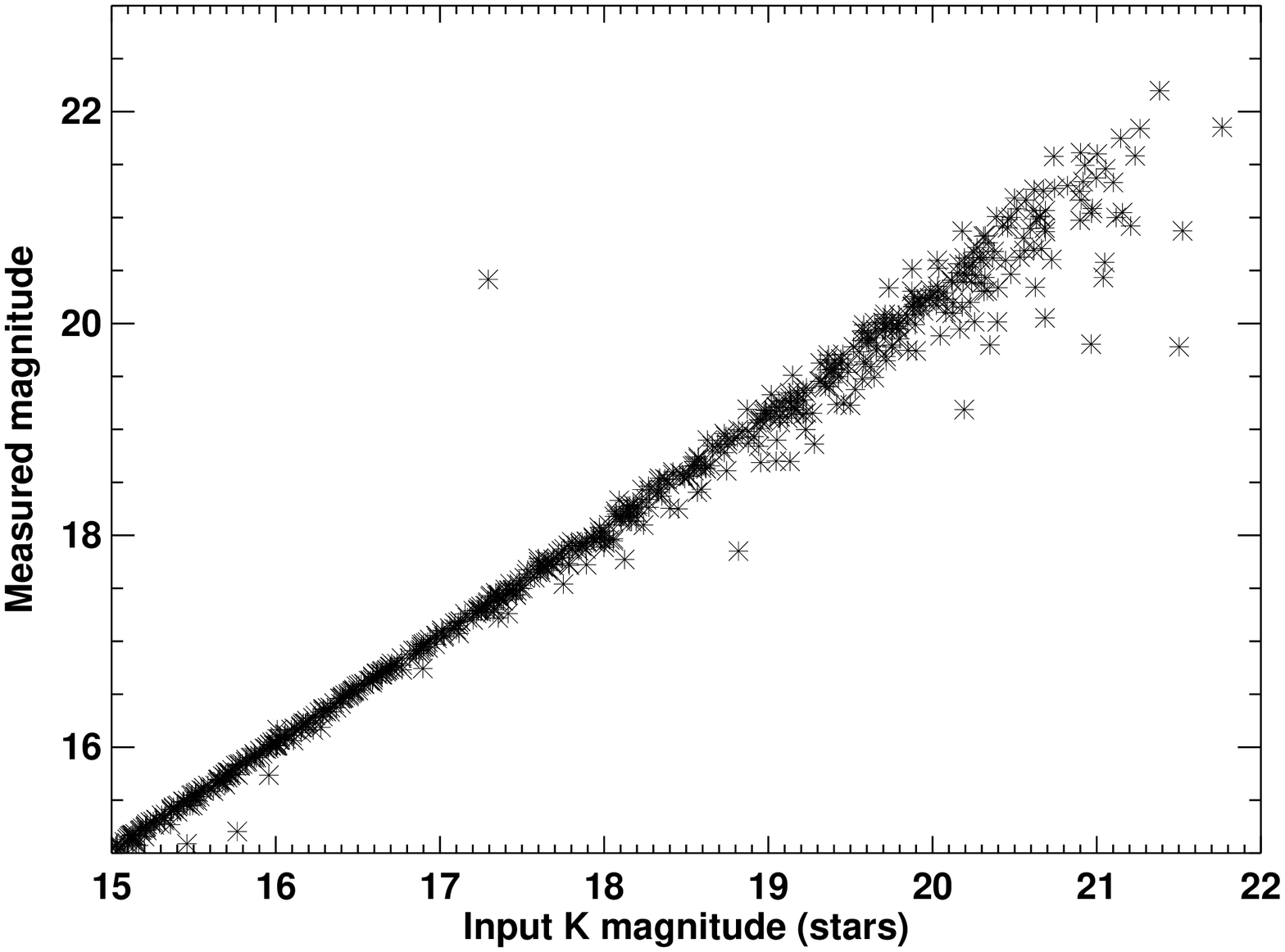,angle=90,width=7.8cm,clip=}
}
\centerline{
\psfig{file=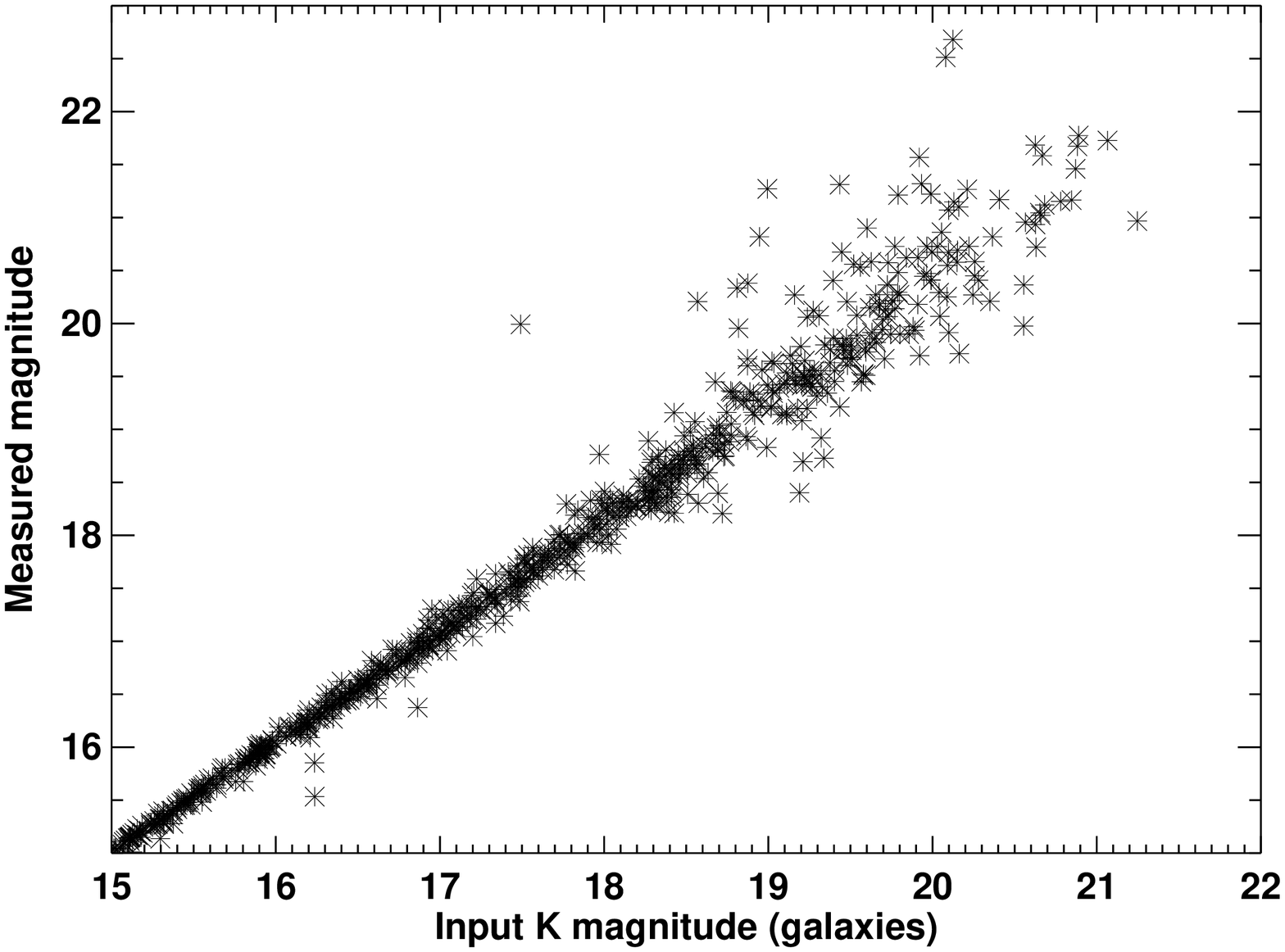,angle=90,width=7.8cm,clip=}
}
\caption{\label{inoutmags} The distribution of measured magnitudes as a
function of input magnitudes for the model stars (upper plot) and galaxies
(lower plot). The plots are for 1000 model objects. No point is plotted
if the model object is not detected by \sextractor.}
\end{figure}

\begin{figure}
\centerline{
\psfig{file=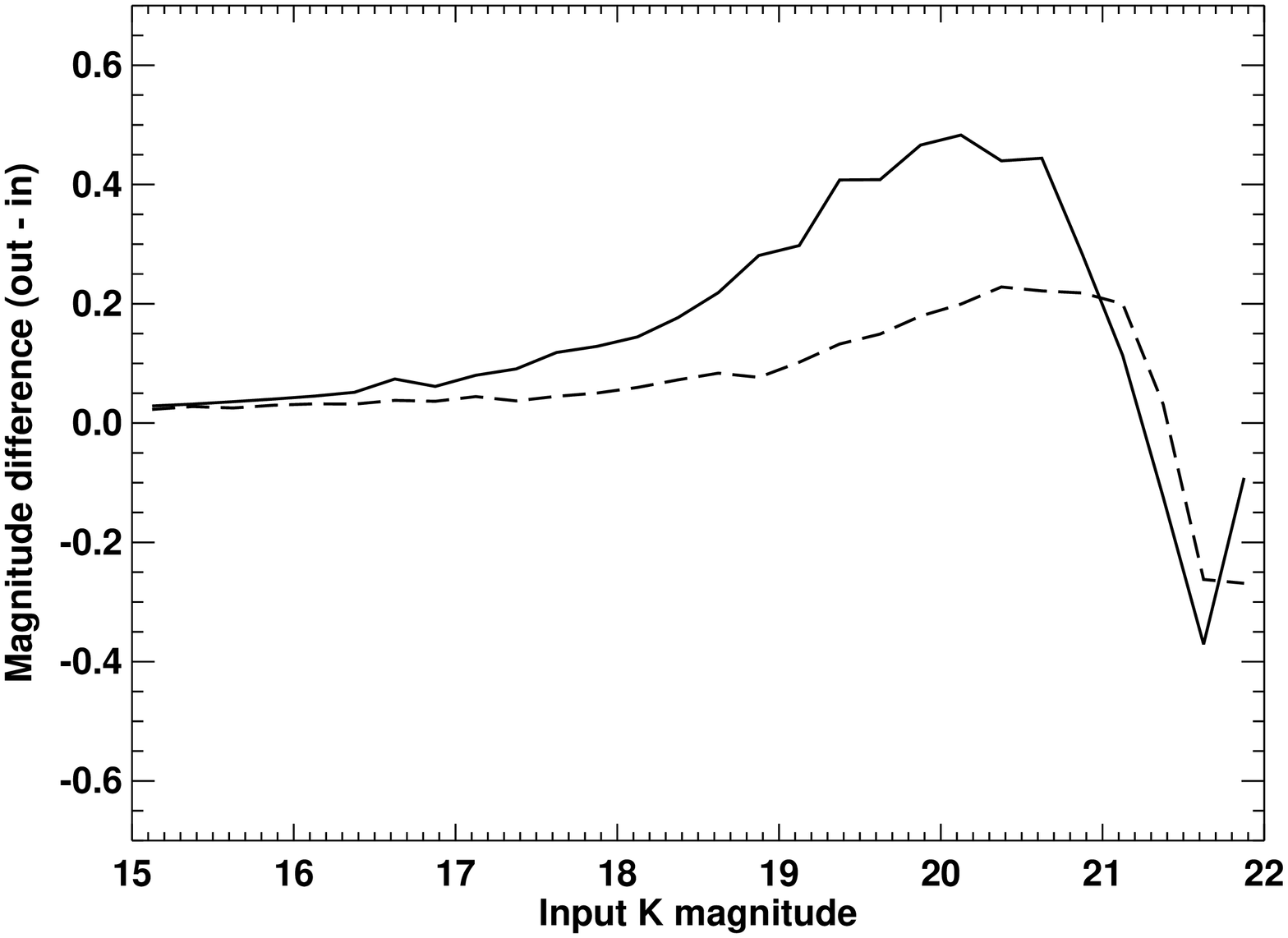,angle=90,width=7.8cm,clip=}
}
\centerline{
\psfig{file=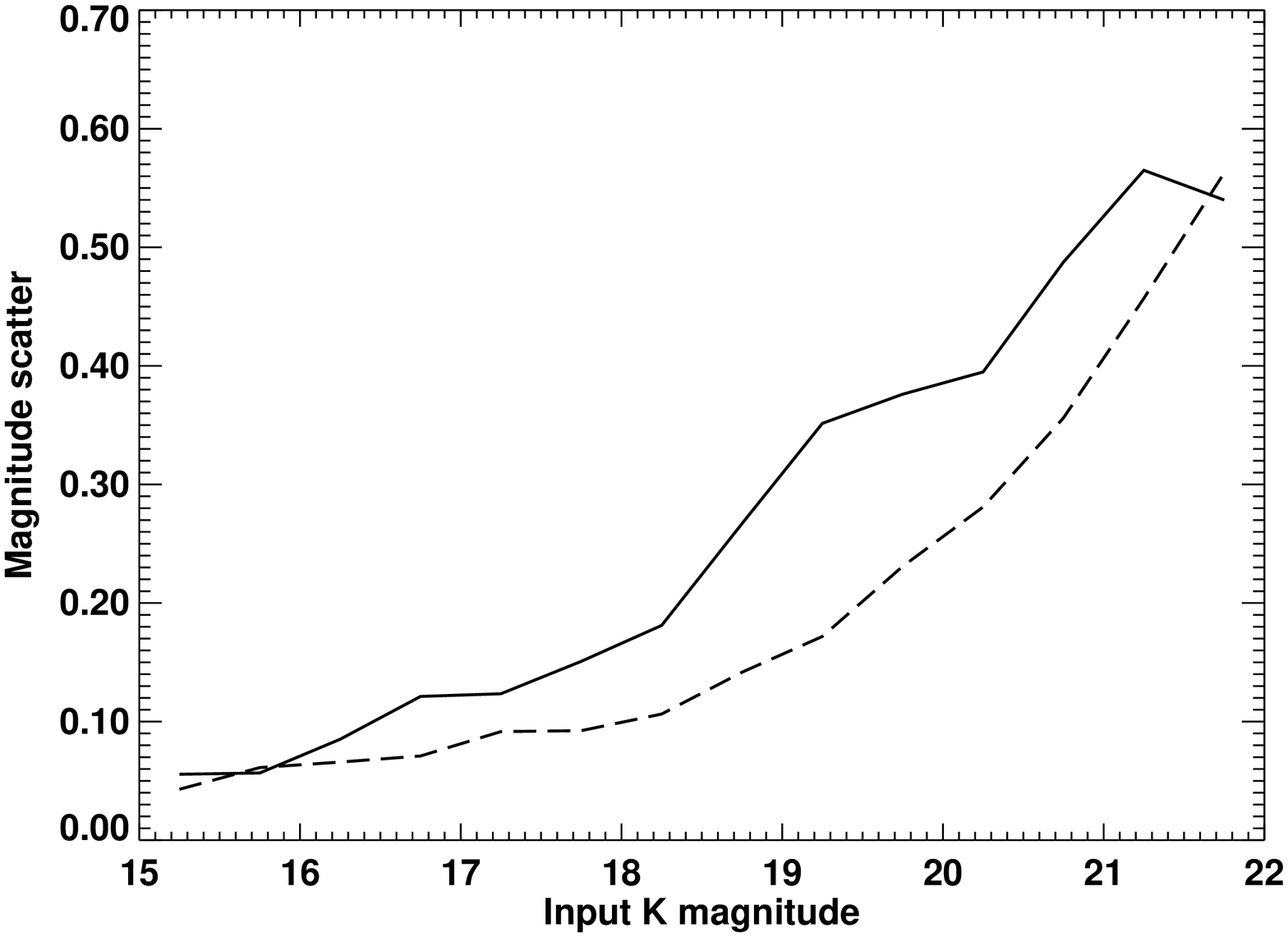,angle=90,width=7.8cm,clip=}
}
\caption{\label{inmagplots} The mean (upper plot) and rms scatter (lower
plot) of the difference between the measured magnitudes and input
magnitudes for the model stars (dashed line) and galaxies (solid line).}
\end{figure}

The input and measured magnitudes of each detected object can also be
compared. Typical results for an individual field are shown in
Figure~\ref{inoutmags}. Note the generally low scatter, particularly at
bright magnitudes, but the occasional source with a large deviation. These
are caused by the proximity of the model object to a bright source in the
field; large errors such as this are avoided in the observed data by
careful examination of the output \sextractor\ catalogues, as discussed
above. The mean difference between the input and measured magnitudes, and
the scatter of the measured magnitudes around this mean, are shown as
functions of input magnitude for both model stars and galaxies in
Figure~\ref{inmagplots}.

\subsection{Multi--colour source extraction}
\label{colsext}

The J--band frames were aligned with the K--band frames by using a number
of objects that appeared unresolved on the HST images. This alignment
often involved a small shift in position of the frames, but generally much
less than a degree of frame rotation. \sextractor\ was then run in its
double image mode using the J and K--band images. In double image mode
\sextractor\ uses one image to detect objects and define the apertures to
be used for flux determination, and then the fluxes and magnitudes are
measured from the second image. In this way J--band magnitudes or upper
limits were determined for the sources detected in the K--band, through
exactly the same apertures. 

For the HST data, \sextractor\ was first run on the HST frames obtained at
the end of the calibration procedure. Then, the four separate WFPC2 frames
were overlaid individually with the K--band data, using between 3 and 15
unresolved objects visible at both wavelengths. In this process the HST
data was re-pixelated to match the UKIRT data, essential for a combined
running of \sextractor\ on the K--band and HST images; the re-pixelated
and aligned HST images were then convolved to the angular resolution of
the K--band data using a Gaussian convolving function. \sextractor\ was
run in double image mode on the convolved HST data, thus measuring
accurate fluxes and magnitudes for the objects through the same apertures
as the K--band data. The HST fluxes were corrected for the small
differences in gain ratios between the different WFPC2 chips and using a
4\% linear correction ramp for the charge transfer efficiency effects
\cite{hol95}. They were also corrected for galactic extinction using the
extinction maps of Burstein and Heiles \shortcite{bur82a}. Finally, for
each object the `stellaricity index' (see Section~\ref{stargalsep}),
used for the separation of stars and galaxies, was replaced by that
calculated in the run of \sextractor\ on the original HST frame: using the
highest angular resolution data provides a much more accurate
determination of this parameter.

For both the J--band and HST frames, only objects which were detected with
fluxes of at least 3$\sigma$ were considered, where $\sigma$ is the
uncertainty on the flux measurement provided by \sextractor; this flux
error estimate includes the uncertainty due to the Poisson nature of the
detected counts and that from the standard deviation of the background
counts. An additional source of flux error arises from the uncertainty in
the subtraction of the background count level as a function of position
across the image. This value was estimated as the product of the area of
the extraction aperture and the rms variation of the subtracted background
flux across the image. This background subtraction error estimate was
combined in quadrature with the flux error given by \sextractor\ to
determine the uncertainties on the magnitudes of the extracted objects.

\subsection{Star--galaxy separation}
\label{stargalsep}

\sextractor\ provides a `stellaricity index' for each object, which is an
indication of the likelihood of an object being a galaxy or a star, based
on a neural network technique \cite{ber96}. In the ideal cases a galaxy
has a stellaricity index of 0.0 and a star has 1.0. In practice, at
faint magnitudes, low signal--to--noise and galaxy sizes smaller than the
seeing lead to an overlap in the calculated stellaricity indices for the
two types of object.

\begin{figure}
\centerline{
\psfig{file=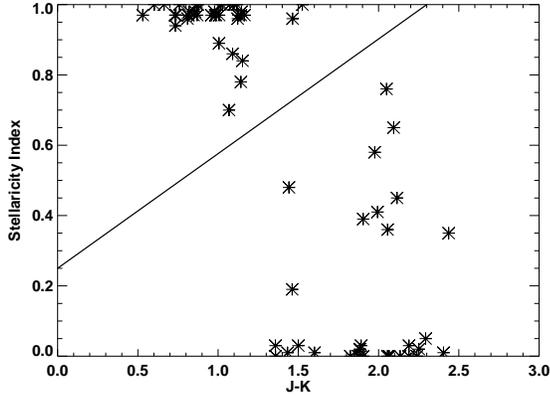,angle=90,width=7.8cm,clip=}
}
\caption{\label{colstel} A plot of stellaricity index against $J-K$
colour: stars generally have bluer colours and higher stellaricity indices
than galaxies, and so combining these two parameters allows fairly
accurate star--galaxy separation.  The stars lie to the upper left of the
line. The field plotted here is 3C22, chosen for this demonstration
because of its high fraction of stars.}
\end{figure}

For sources which were present on the HST frames (typically only about
80\% of objects, the precise percentage depending upon the sky rotation
angle at which the HST frame was taken), the stellaricity index for each
object determined from the un-convolved HST data was adopted: the high
resolution of these images meant that, apart from the very reddest
objects, star--galaxy separation\footnote{Of course, star--galaxy
separation using the resolved or unresolved nature of sources means that
any quasars will be placed into the star category, but their contribution
is negligible.} was relatively unambiguous right down to the completeness
limit of the K--band frames. For the objects for which no HST data was
available, stellaricity indices were taken from the K--band data. At
magnitudes $K \lta 17.5$, star--galaxy separation could be carried out
directly from these stellaricity values. At fainter magnitudes it was
still possible to provide a fairly accurate segregation of galaxies from
stars by combining the stellaricity index with the $J-K$ (or F814W$-K$)
colour of the object, as shown in Figure~\ref{colstel}: the bluer average
colour of stars can be used to separate out the stars and galaxies in the
ambiguous range of stellaricity indices from about 0.5 to 0.9.

\begin{figure}
\centerline{
\psfig{file=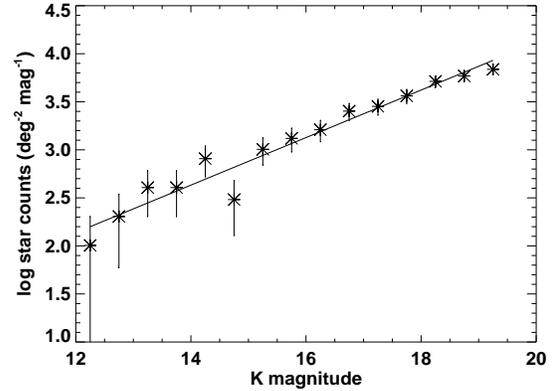,angle=90,width=7.8cm,clip=}
}
\caption{\label{starcnts} The differential star number counts in the
K--band, in $\Delta K = 0.5$ magnitude intervals, fit using a single power
law distribution.}
\end{figure}

Figure~\ref{starcnts} shows the total star counts as a function of
magnitude derived from all of the frames, fit with the function $\log N(K)
= 0.247 K - 0.830$. This good fit to the differential star counts using a
single power--law distribution clearly demonstrates that the star--galaxy
separation is working well.

\section{Galaxy Counts}
\label{galcounts}

Using the differences between true and measured magnitudes determined for
the simulated galaxies in Section~\ref{sextract} (and Figure~3), the
measured `total' magnitudes of the galaxies were converted to true `total'
magnitudes, and these were binned in 0.5 magnitude bins over the magnitude
range $14 < K < 20$ to determine the raw galaxy counts, $n_{\rm raw}$.
These were corrected for completeness using the observed mean completeness
fraction as a function of magnitude from the simulations
(Figure~\ref{compplot}). A possible source of systematic error in the
number counts at faint magnitudes must also be taken into account: the
increased scatter in the photometric measurements at faint magnitudes, and
the fact that there are more faint galaxies than bright galaxies, mean
that it is more likely for faint galaxies to be brightened above the
$K=20$ limit (or be brightened and move up one magnitude bin) than for
brighter galaxies to appear erroneously in a fainter bin or fall out of
the catalogue completely.  This leads to an apparent increase in the
number counts at the faintest magnitudes. The results from the model
galaxy simulations were used to correct for this effect: the required
corrections were smaller than 10\% in all cases. Combining these effects
produced corrected galaxy counts ($n_{\rm c}$), and scaling by the
observed sky area, produced final counts per magnitude per square degree,
$N_{\rm c}$.

Number counts were also derived using the same technique considering just
the five fields with deeper images (see Section~\ref{dataset}). This
provided determinations of galaxy counts for $17.5 < K < 20.5$ which may
be more reliable at the fainter magnitudes due to smaller completeness
corrections. 

\begin{table}
\caption{\label{galcntstab} K--band galaxy counts as a function of
magnitude. The columns give the raw counts ($n_{\rm raw}$), the counts
corrected for completeness and magnitude biasing effects ($n_{\rm c}$),
the corrected counts per square degree per unit magnitude ($N_{\rm c}$),
the error on this, and the mean literature counts per square degree per
unit magnitude from the data shown in Figure~\ref{galcntsplot} ($N_{\rm
lit}$).}
\begin{tabular}{crrrrr}
$K$&$n_{\rm raw}$&$n_{\rm c}$ &$N_{\rm c}$$^*$ &$\delta N_{\rm c}$$^*$
&$N_{\rm lit}$$^*$ \\
\multicolumn{2}{l}{All fields} \\
14.0 -- 14.5  &   1   &    1.0   &    101  &    101  &     48 \\
14.5 -- 15.0  &   2   &    2.0   &    202  &    143  &     85 \\
15.0 -- 15.5  &   3   &    3.0   &    304  &    175  &    218 \\
15.5 -- 16.0  &  10   &   10.1   &   1021  &    320  &    620 \\
16.0 -- 16.5  &  18   &   18.2   &   1840  &    430  &    960 \\
16.5 -- 17.0  &  36   &   36.4   &   3680  &    610  &   1530 \\
17.0 -- 17.5  &  63   &   64.3   &   6500  &    830  &   2530 \\
17.5 -- 18.0  &  93   &   90.8   &   9180  &   1010  &   5540 \\
18.0 -- 18.5  & 125   &  123.7   &  12500  &   1240  &   9470 \\
18.5 -- 19.0  & 189   &  203.0   &  20500  &   1770  &  10500 \\
19.0 -- 19.5  & 174   &  208.9   &  21100  &   2450  &  11600 \\
19.5 -- 20.0  & 168   &  306.7   &  31000  &   4700  &  17200 \\
\multicolumn{2}{l}{Deep fields only} \\	    
17.5 -- 18.0  &  18   &   18.4   &   9340  &   2190  &   5540 \\
18.0 -- 18.5  &  23   &   21.6   &  10980  &   2510  &   9470 \\
18.5 -- 19.0  &  43   &   44.6   &  22700  &   3610  &  10500 \\
19.0 -- 19.5  &  46   &   47.9   &  24300  &   4080  &  11600 \\
19.5 -- 20.0  &  59   &   83.1   &  42200  &   6600  &  17200 \\
20.0 -- 20.5  &  54   &~~114.6   &~~58200  &~~10900  &~~22700 \\
\\
\multicolumn{6}{l}{$^*$: counts per magnitude per square degree.}
\end{tabular}
\end{table}

\begin{figure}
\centerline{
\psfig{file=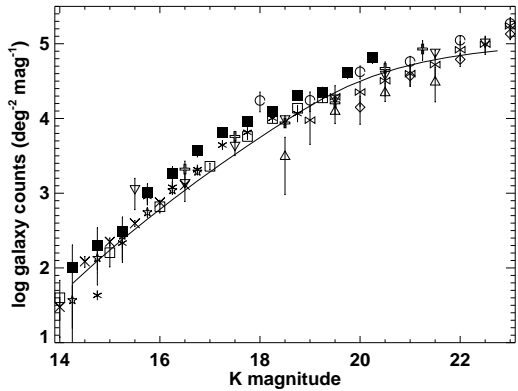,angle=90,width=7.8cm,clip=}
}
\caption{\label{galcntsplot} The K--band galaxy number counts per
magnitude per square degree, in $\Delta K = 0.5$ magnitude intervals. The
solid symbols represent the data presented in this paper and tabulated in
Table~\ref{galcntstab}; for $K < 19$ the values plotted are those from
`all fields' and for $K>19$ they are those from the 5 deep fields
only. The other symbols represent data from the literature: open squares
and (point-up) triangles are respectively the bright and faint samples of
Minezaki \etal\ (1998); diamonds are from Djorgovski \etal\ (1995);
circles are from Moustakas \etal\ (1997); x's are from Szokoly \etal\
(1998); asterisks, inverted triangles and stars are respectively the
Hawaii Medium Deep Survey, the Hawaii Deep Survey and the Hawaii Medium
Wide Survey presented by Gardner, Cowie and Wainscoat (1993); open crosses
are from McLeod \etal\ (1995); bow-ties ($\bowtie$) are from Bershady,
Lowenthal and Koo (1998). The solid line represents a passive luminosity
evolution model, discussed in Section~\ref{spatcorr}.}
\end{figure}
\nocite{min98,djo95,mou97,szo98,gar93,mcl95,ber98}

These galaxy counts are tabulated in Table~\ref{galcntstab} and plotted in
Figure~\ref{galcntsplot}. The error on the number counts in each bin was
calculated from two factors: the Poissonian error on the raw galaxy
counts, and an error in the completeness correction which is generously
assumed to be 30\% of the number of counts added in the correction.  The
number counts are compared on the plot with counts from various K--band
field surveys in the literature. The final column in
Table~\ref{galcntstab} shows the mean galaxy counts per magnitude bin
determined from these literature field surveys.

The galaxy counts derived in this paper show an excess of counts relative
to the literature counts, which for magnitudes $K > 15.5$ is at greater
than the $1\sigma$ significance level. For comparison, the K--magnitudes
of the radio galaxies themselves span the range $15.4 < K <18.0$. The
integrated excess counts over the magnitude range $15.5 < K < 20$
corresponds to, on average, 11 galaxies per field. Unfortunately, as
described in Section~\ref{dataset}, the original goals of these
observations was to carry out detailed studies of the radio source host
galaxies, and so no blank sky frames were taken. Therefore no comparison
can be made between radio galaxy and blank sky frames to determine whether
the excess counts found here are associated with structures surrounding
the radio galaxies or whether there is some systematic offset between the
counts calculated here and the literature counts.

We believe that the majority of the excess counts are associated with the
presence of the radio galaxies, for a number of reasons. First, Roche,
Eales and Hippelein \shortcite{roc98a} found a similar excess in the
K--band galaxy counts in a study of the fields surrounding 6C radio
galaxies at similar redshifts. Second, as reviewed in the introduction,
there are numerous lines of evidence suggesting that there are clusters
around at least some of these objects. Third, later in this paper it will
be shown that cross--correlation analyses and colour--magnitude relations
indicate an overdensity of galaxies comparable to that observed.

\section{Cross--correlation analyses}
\label{crosscor}

\subsection{Angular cross-correlation estimators}
\label{angcross}

The clustering of galaxies around the radio galaxies can be investigated
using the angular cross--correlation function $w(\theta)$, which is
defined from the probability of finding two sources in areas
$\delta\Omega_1$ and $\delta\Omega_2$ separated by a distance $\theta$:

\begin{displaymath}
\delta P = N^2 [1 + w(\theta)] \delta\Omega_1 \delta\Omega_2
\end{displaymath}

\noindent where $N$ is the mean surface density of sources on the
sky. From this, it follows that for a given survey:

\begin{equation}
DD(\theta) = \frac{1}{2} n_{\rm D} (n_{\rm D}-1) [1 + w(\theta)]
\frac{\langle\delta\Omega_{\rm D}(\theta)\rangle}{\Omega},
\label{dddef}
\end{equation}

\noindent where $DD(\theta)$ is the number of data--data pairs with
angular separation of between $\theta$ and $\theta + \delta\theta$,
$n_{\rm D}$ is the total number of sources in data catalogue, $\Omega$ is the
total angular area of sky sampled and $\langle\delta\Omega_{\rm
D}(\theta)\rangle$ is the mean angular area of sky accessible at a
distance $\theta$ to $\theta + \delta\theta$ around the data
points. Clearly $\langle\delta\Omega_{\rm D}(\theta)\rangle$ is extremely
difficult to calculate due to boundary effects; various estimators for
$w(\theta)$ have therefore been derived using comparisons between the data
points and catalogues of randomly distributed points (e.g. see Cress
\etal\ 1996 for a discussion).  For the estimator adopted in this paper,
it is considered that if $n_{\rm R}$ random points are added to the image
(where $n_{\rm R} \gg n_{\rm D}$ to minimise errors introduced by the
random catalogue) then the number $DR(\theta)$ of data--random pairs
between $\theta$ and $\theta + \delta\theta$ will be given by:

\begin{equation}
DR(\theta) = n_{\rm D} n_{\rm R} \frac{\langle\delta\Omega_{\rm
D}(\theta)\rangle}{\Omega}
\label{drdef}
\end{equation}

\noindent Combining equations~\ref{dddef} and~\ref{drdef} gives the following
estimator for $w(\theta)$:

\begin{equation}
w(\theta) = \frac{2 n_{\rm R}}{(n_{\rm D}-1)}
\frac{DD(\theta)}{DR(\theta)} - 1 
\label{wthetadef}
\end{equation}

In addition, it is possible to consider the angular cross--correlation
function just around an individual source, in this case the radio galaxy.
For this, the parameters become $RGD(\theta) = (n_{\rm D}-1) [1 + w_{\rm
rg}(\theta)] \langle\delta\Omega_{\rm RG}(\theta)\rangle / \Omega$ and
$RGR(\theta) = n_{\rm R} \langle\delta\Omega_{\rm RG}(\theta)\rangle /
\Omega$, where $RGD(\theta)$ and $RGR(\theta)$ are respectively the number
of radio galaxy--data and radio galaxy--random pairs between $\theta$ and
$\theta + \delta\theta$, $w_{\rm rg}(\theta)$ is the angular
cross--correlation function for galaxies around the radio galaxy, and
$\langle\delta\Omega_{\rm RG}(\theta)\rangle$ is the angular area of sky
accessible at a distance $\theta$ to $\theta + \delta\theta$ from the
radio galaxy. These equations combine to give:

\begin{equation}
w_{\rm rg}(\theta) = \frac{n_{\rm R}}{n_{\rm D}}
\frac{RGD(\theta)}{RGR(\theta)} - 1 
\label{wthetargdef}
\end{equation}

\subsection{Calculating $w(\theta)$ and $w_{\rm rg}(\theta)$}
\label{wthetacalc}
   
One problem with estimating $w(\theta)$ and $w_{\rm rg}(\theta)$ from the
current data is that, due to the K--band dithering technique, the
detection limit is a function of position across each image. More objects
are detected in the central regions of the image, which would produce a
spurious peak in the angular cross--correlation statistics. This
non-uniform noise level can be accounted for using the simulations which
were discussed in Section~\ref{sextract}. From these simulations, for each
field the completeness fraction can be calculated as a function of both
magnitude and position, and then applied to the random catalogue in the
following way.

\noindent {\it Step 1:} Excluding those objects determined to be stars,
$DD(\theta)$, $RGD(\theta)$, and $n_{\rm D}$ were determined for each
frame down to a chosen limiting magnitude $K_{\rm lim}$, for a set of bins
in $\theta$.

\noindent {\it Step 2:} 25000 random objects were each assigned a random
position on the image, and a magnitude $K \le K_{\rm lim}$ drawn at random
from a distribution matching the observed number counts
(Figure~\ref{galcntsplot}).

\noindent {\it Step 3:} For each random object, the completeness fraction
of objects of that magnitude and position (averaged in 50 by 50 pixel
bins) was determined from the simulations. The object was accepted or
rejected at random with a likelihood of acceptance based upon the derived
completeness fraction.

\noindent {\it Step 4:} For those objects which remained in the catalogue,
$DR(\theta)$, $RGR(\theta)$, and $n_{\rm R}$ were calculated. Hence
$w(\theta)$ and $w_{\rm rg}(\theta)$ were derived.

\begin{figure*}
\centerline{
\psfig{file=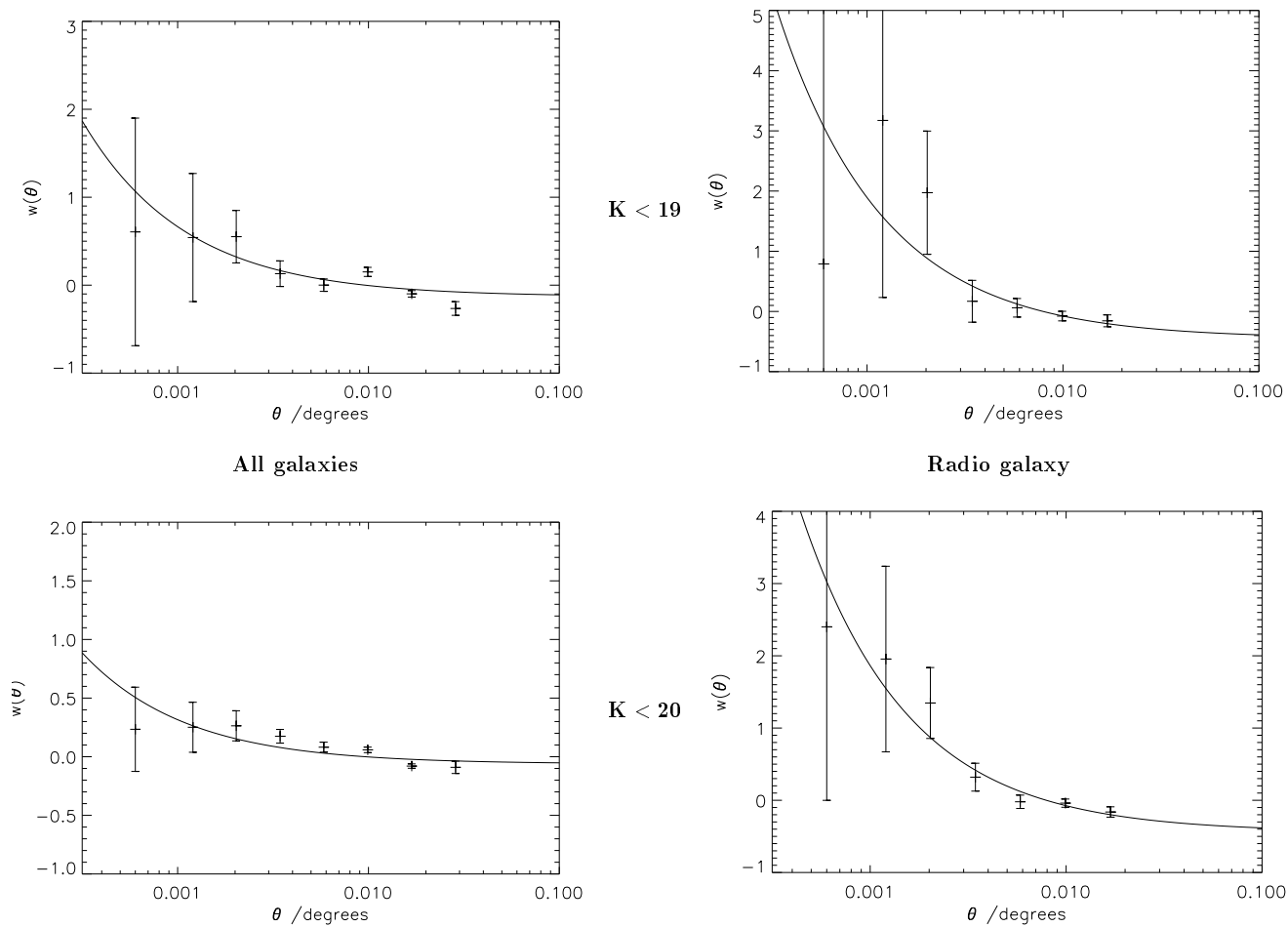,angle=-90,width=\textwidth,clip=} 
}
\caption{\label{wthetaplot} The angular cross--correlation function for
all galaxy--galaxy pairs (left) and for just the radio galaxy--galaxy
pairs (right) averaged over all of the UKIRT K--band images. The upper
plots include all galaxies with $K<19$, to which level the observations
are almost complete; the lower plots include galaxies down to $K<20$, at
which level there is a significant completeness correction, but the number
statistics are increased. The plots are fit with functions of the form
$w(\theta) = A\theta^{-0.8} - C$, where $C$ is the integral constraint and
has a value of $C_{\ggal} = 41.1 A_{\ggal}$ for all galaxy--galaxy pairs,
and $C_{\rggal} = 48.0 A_{\rggal}$ for just radio galaxy--galaxy pairs
(see text).}
\end{figure*}

This process was repeated for all images and for two different limiting
magnitudes, $K_{\rm lim}=19$ and $K_{\rm lim}=20$. $K_{\rm lim}=19$
corresponds to the limit at which the data are still almost 100\%
complete, whilst $K_{\rm lim}=20$ requires a significant completeness
correction but provides improved statistics for the galaxy counts. The
results were averaged over all 28 fields; investigating the values for
individual fields was not attempted since for these small fields of view
the measured amplitudes of single images are too strongly affected by
variations in the background (and foreground) counts to provide results
with a high statistical significance; only by combining the data of the 28
fields is a sufficiently robust measurement obtained.  An estimation for
the uncertainty in the value of the combined $w(\theta)$ in each bin was
provided by the scatter in the values between the different fields. The
results are shown in Figure~\ref{wthetaplot}.

$w(\theta)$ is usually assumed to have a power--law form; $w(\theta) = A
(\theta / {\rm deg})^{-\delta}$. If so, the observed $w(\theta)$ will
follow a form $w(\theta) = A (\theta / {\rm deg})^{-\delta} - C$, where $C$
is known as the integral constraint and arises from the finite size of the
field of view. Its value can be estimated by integrating $w(\theta)$ over
the area of each field, and for our data corresponds to $C_{\ggal} = 41.1
A_{\ggal}$ and $C_{\rggal} = 48.0 A_{\rggal}$ for, respectively, all
galaxy-galaxy pairs and just radio galaxy-galaxy pairs. It is not possible
to determine both the amplitude and the slope of the fit from the current
data, and so the canonical value of $\delta = 0.8$ (which also seems to be
appropriate at high redshifts; Giavalisco \etal\ 1998)\nocite{gia98} has
been adopted. The observed data in Figure~\ref{wthetaplot} has been fit
with functions of this form, and the resulting angular cross--correlation
amplitudes are provided in Table~\ref{angres}.

\begin{table}
\caption{\label{angres} Amplitudes of the angular and spatial
cross--correlation functions for the fields surrounding the 3CR
galaxies. Both the amplitudes for all galaxy--galaxy pairs and those
considering only radio galaxy--galaxy pairs are given, down to two
limiting magnitudes. For radio galaxy--galaxy pairs these are converted 
to spatial cross--correlation amplitudes as described in
Section~\ref{spatcorr}.} 
\begin{tabular}{cccccc}
Sources & $K_{\rm lim}$  & $A$ & $\Delta A$ & $B$ &$\Delta B$\\
\\
All g-g pairs   & $K<19$ & 0.0031 & 0.0012  \\      
                & $K<20$ & 0.0015 & 0.0004  \\    
Only rg-g pairs & $K<19$ & 0.0092 & 0.0037  & 600 & 240 \\     
                & $K<20$ & 0.0093 & 0.0021  & 510 & 120 \\
\end{tabular}
\end{table}

The results at $K<19$ and $K<20$ are in approximate agreement. For the
galaxy--galaxy pairs there may be a decrease of $A_{\ggal}$ as fainter
magnitude limits are used (as expected, e.g. see Roche \etal\
1998)\nocite{roc98a}, but the errors are too large to determine this with
any degree of confidence. The galaxy--galaxy amplitudes are similar to
those derived by Roche \etal\ to the same magnitude limits in the fields
of radio galaxies at $z \sim 0.75$. A more obvious feature is that at both
magnitude limits the cross--correlation amplitude around the radio galaxy
is significantly larger ($A_{\rggal} \gg A_{\ggal}$).
Figure~\ref{wthetaplot} shows that much signal originates at small angular
separations ($\theta \lta 10''$) and so this cannot be related to any
problems with completeness in the outer regions of the frames; the
similarity of the results for $K<19$ and $K<20$ also demonstrates this.

\subsection{The spatial cross--correlation amplitude}
\label{spatcorr}

As has been described by many authors (e.g. Longair \& Seldner 1979,
Prestage and Peacock 1988)\nocite{lon79b,pre88} it is possible to convert
from an angular cross--correlation amplitude to a spatial
cross--correlation amplitude if the galaxy luminosity function is
known. The spatial cross--correlation function is usually assumed to have
a power--law form:

\begin{displaymath}
\xi(r) = B_{\rggal} \left(\frac{r}{{\rm Mpc}}\right)^{-\gamma}
\end{displaymath}

\noindent where the power--law slope is related to the slope of the
angular cross--correlation function by $\gamma = \delta + 1$ ($=$1.8 for
the $\delta = 0.8$ adopted here). The spatial cross--correlation
amplitude, $B_{\rggal}$, is then related to that of the angular
cross-correlation by $A_{\rggal} = H(z) B_{\rggal}$ (see Longair \& Seldner
\shortcite{lon79b} for a full derivation) where:

\begin{equation}
H(z) = \frac{I_{\gamma}}{N_{\rm g}} \left(\frac{D(z)}{1+z}\right)^{3-\gamma}
\phi(m_{\rm 0},z) \left(\pi / 180\right)^{-(\gamma-1)}
\end{equation}

\noindent Here, $I_{\gamma}$ is a definite integral which for $\gamma =
1.8$ has a value of 3.8, $N_{\rm g}$ is the measured sky density of
objects above the magnitude limit (per steradian), $D(z)$ is the proper
distance to a source at redshift $z$, $\phi(m_{\rm 0},z)$ is the number of
galaxies per unit comoving volume which at redshift $z$ are more luminous
than apparent magnitude $m_{\rm 0}$, and the factor $\left(\pi /
180\right)^{-(\gamma-1)}$ is required to convert $A_{\rggal}$ from degrees
to radians. The value of $H(z)$ is only a very weak function of $\gamma$
provided $\gamma$ is of order 2 \cite{pre88}, and so the fixing of
$\gamma$ at 1.8 will introduce no significant errors.

$\phi(m_{\rm 0},z)$ requires assumptions for both the K--correction of
galaxies and the local luminosity function, as well as being cosmology
dependent, and so this conversion will always be somewhat uncertain. In
this paper a Schechter \shortcite{sch76} form is adopted for the local
luminosity function, that is:

\begin{displaymath}
\phi(L) {\rm d}L = \phi^* \left(\frac{L}{L^*}\right)^{\alpha}
\exp\left(\frac{-L~}{L^*}\right){\rm d}\left(\frac{L}{L^*}\right)  
\end{displaymath}

\noindent where $\phi(L) {\rm d}L$ is the number of galaxies per comoving
cubic Mpc with luminosity $L$ to $L + {\rm d}L$, $\phi^*$ is the density
normalization factor, $L^*$ is the characteristic luminosity and $\alpha$
is the faint--end slope. A pure luminosity evolution model is also adopted;
that is, values for $\phi^*$, $\alpha$ and $L^*(z=0)$ are taken, and the
only evolution of this function is then evolution of the value of $L^*$
with redshift in accordance with passive evolution predictions.  As
discussed in the introduction, passive evolution models provide a good
description of cluster galaxy properties back to $z \sim 1$, and they can
also provide a reasonable fit to the observed K--band field number counts
(see below). These models are, however, undoubtedly a simplification in
view of hierarchical galaxy formation theories. In hierarchical models,
fewer luminous galaxies are predicted to exist at high redshifts (e.g. see
Figure~4 of Kauffmann and Charlot 1998)\nocite{kau98a}, resulting in the
values of $\phi(m_{\rm 0},z)$ and consequently $H(z)$ being lower, and
hence $B_{\rggal}$ will be increased. The pure passive luminosity approach
therefore provides a conservative lower estimate for the spatial
cross--correlation amplitude around high redshift radio galaxies.

For the passive evolutionary models, the galaxy population was split into
four different galaxy types: ellipticals and S0's (E's), Sa and Sb types
(Sab's), Sc types (Sc's) and Sd types and irregulars (Sdm's). Galaxies of
these types were built up using the Bruzual and Charlot \shortcite{bru93}
stellar synthesis codes (1996 version), assuming a Scalo initial mass
function, solar metallicity, a formation redshift $z_{\rm f} = 10$, and
four different star formation histories (cf. Guiderdoni and
Rocca--Volmerange 1990)\nocite{gui90}. The E's were assumed to form their
stars in a rapid early burst with the star formation decreasing
exponentially on a $\tau_{\rm sfr} = 0.5$\,Gyr timescale. The Sab's had
50\% of their stars in a bulge component formed in the same manner as the
E's, and the remaining 50\% in a disk--like component with a much longer
star formation timescale ($\tau_{\rm sfr} = 6$\,Gyr). The Sc's were
modelled using just a single long star formation timescale ($\tau_{\rm
sfr} = 8$\,Gyr), and the Sdm's using a still longer timescale ($\tau_{\rm
sfr} = 50$\,Gyr) to produce significant star formation at the current
epoch. These star formation histories approximately reproduce the colours
of the various morphological types at the current epoch. It was assumed
that all of these four morphological types displayed the same luminosity
function (that is, $M^*$ and $\alpha$), with the weightings of the
different types (ie. the relative contributions to $\phi^*$) being 30\%,
30\%, 30\% and 10\% respectively. In fact, at wavelengths as long as the
K--band, the K and evolutionary corrections are small and relatively
independent of morphological type, and so the results are not strongly
dependent upon any of these assumptions. Changing the values adopted above
by a sufficiently large amount that they produce unacceptable
distributions of galaxy colours at the current epoch produces only 10-20\%
changes in the final K--band luminosity function.

Local K--band luminosity functions have been derived by Gardner \etal\
\shortcite{gar93}, by Mobasher, Sharples and Ellis \shortcite{mob93} and
by Loveday \shortcite{lov00}. All three of these datasets are
approximately consistent with a characteristic absolute magnitude
$M^*_{\rm K} = -25.1$ and a slope $\alpha = -1.0$. To investigate whether
the combination of this luminosity function and the assumption of passive
evolution models produces acceptable results, and to calculate the
appropriate value of $\phi^*$, the K--band number counts as a function of
apparent magnitude that would be expected in this model were derived. The
value of $\phi^*$ was adjusted to provide a good fit to the observed data
points. The resulting fit, for a value $\phi^* \approx 0.004$, is shown as
the solid line in Figure~\ref{galcntsplot}, and demonstrates that this
simple model can provide a reasonable fit to the observed number counts.

With this model luminosity function, it is possible to calculate
$\phi(m_{\rm 0},z)$ and, using the values of $N_{\rm g}$ from
Table~\ref{galcntstab} ($N_{\rm g} = 9.2 \times 10^7 {\rm sr}^{-1}$ for
$K<19$ and $2.0\times 10^8 {\rm sr}^{-1}$ for $K<20$), hence to determine
$H(z)$. For $K<19$, $H(z)$ ranges from $15.4 \times 10^{-7}$ to $1.8
\times 10^{-7}$ over the redshift range 0.6 to 1.8. The mean value of $1 /
H(z)$ averaged over all 28 radio source redshifts is $\overline{1 / H(z)}
\approx 1.7 \times 10^6$, and using this value to convert from $A_{\rggal}$
to $B_{\rggal}$ gives $B_{\rggal} = 600 \pm 240$. For $K<20$ the corresponding
range of $H(z)$ is from $12.2 \times 10^{-7}$ to $3.1 \times 10^{-7}$,
with a mean value of $\overline{1 / H(z)} \approx 1.4\times 10^6$,
corresponding to $B_{\rggal} = 510 \pm 120$.

These values can be interpreted physically by comparing with the
equivalent values for Abell clusters calculated between the central galaxy
and the surrounding galaxies ($B_{\cgal}$). This has been calculated
independently as a function of Abell cluster richness by a number of
authors \cite{pre88,hil91,and94,yee99}. Converting their values to $H_0 =
50$\,km\,s$^{-1}$\,Mpc$^{-1}$ and $\gamma = 1.8$ they are all
approximately consistent with each other, and average to $B_{\cgal}
\approx 350$ for Abell class 0 and $B_{\cgal} \approx 710$ for Abell class
1. The environments surrounding the redshift one radio galaxies are
therefore comparable, on average, to those of clusters of between Abell
Classes 0 and 1 richness.

Hill and Lilly \shortcite{hil91} further showed that there is a
correlation between the value of $B_{\cgal}$ and the parameter $N_{0.5}$,
where $N_{0.5}$ is an Abell--type measurement defined as the net excess
number of galaxies within a radius of 0.5\,Mpc of the central galaxy with
magnitudes between $m_1$ and $m_1 + 3$, $m_1$ being the magnitude of the
central galaxy. Using a larger dataset, Wold \etal\ \shortcite{wol00}
calibrated this relation as $B_{\cgal} = (37.8 \pm 10.9) N_{0.5}$. The
average $B_{\rggal}$ value for $K<20$ then implies that an average net
excess of 13.5 galaxies should be found around each radio galaxy within a
radius of 0.5\,Mpc and with a magnitude down to three magnitudes fainter
than the radio galaxy. The data presented in this paper cover about 80\%
of this sky area and, since the radio galaxies have typical magnitudes of
$K \sim 17$, the galaxy counts to $K=20$ do sample approximately 3
magnitudes below the radio galaxy. Therefore, the net excess counts of 11
galaxies per field down to $K=20$ (see Section~\ref{galcounts}) is fully
consistent with the determined value of $B_{\rggal}$.

\section{Colour--magnitude and colour--colour relations}
\label{colmagsect}

\subsection{Near--infrared colour--magnitude relations}

For each field the magnitudes of all galaxies were determined through all
of the different filters, as described in Section~\ref{colsext}. After
correction for galactic reddening, these were used to construct
colour--magnitude relations for each field. The near--infrared $J-K$ vs
$K$ relation for each field (if $J$--band data were not taken, the
longest wavelength HST filter available was used instead) are shown in
Figure~\ref{colmagfigs} in order of ascending redshift. The stars (open
diamonds) and galaxies (asterisks) have been separated on these diagrams
using the technique described in Section~\ref{stargalsep}, and the radio
galaxy is plotted using a large symbol.  The uncertainties on the measured
colours are not shown to avoid cluttering up the diagrams; at bright
magnitudes ($K \lta 17$) they are dominated by calibration uncertainties
($\lta 0.1$ mags), but increase to about 0.3 magnitudes by $K \sim 20$.

\begin{figure*}
\centerline{
\psfig{file=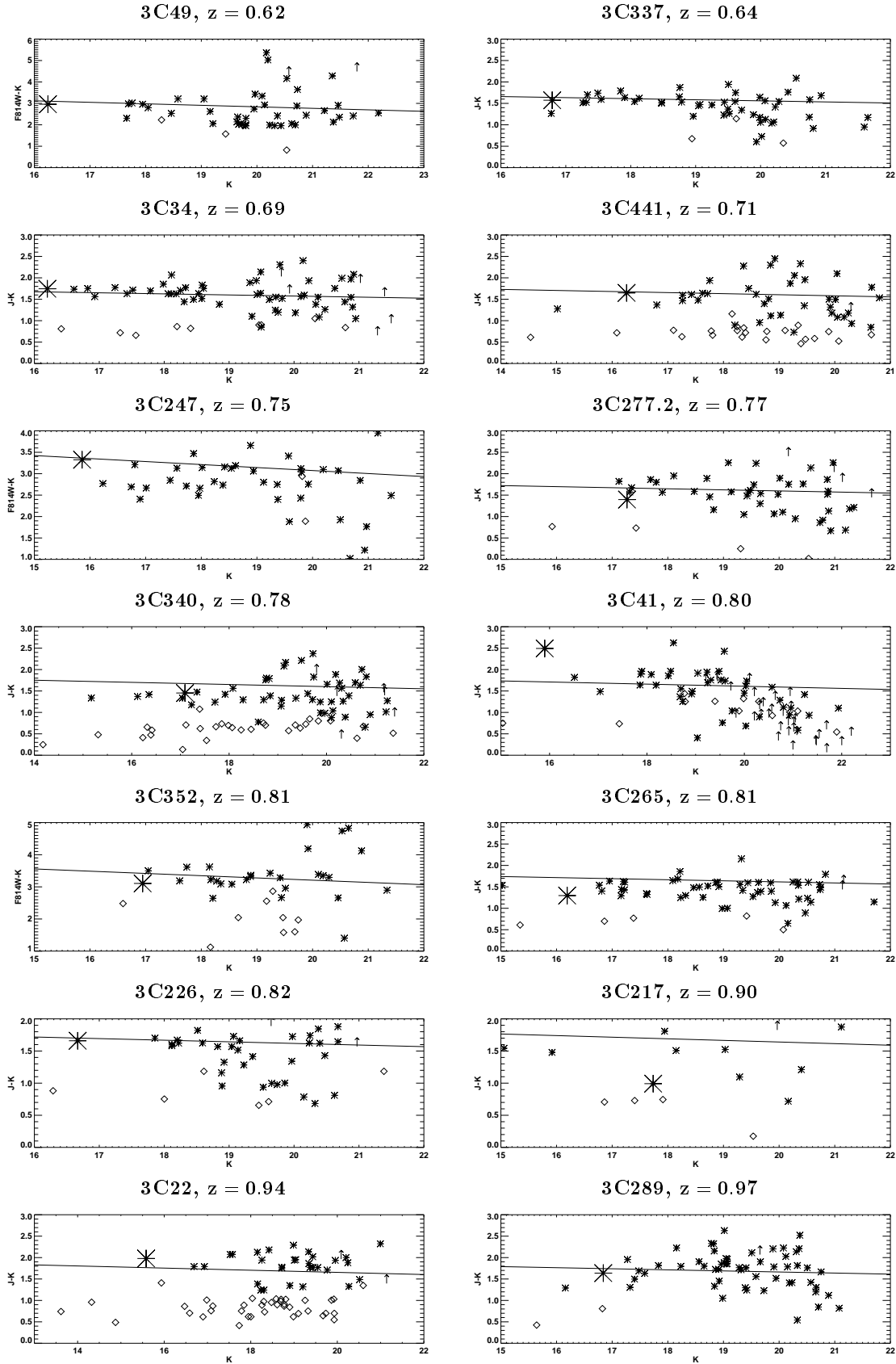,width=14.9cm,clip=}  
}
\caption{\label{redcolfigs} Near--infrared colour--magnitude relations for
the 3CR radio galaxies. The fields and redshifts are labelled above each
diagrams. Galaxies are represented by asterisks and stars by open
diamonds. Upper limits are plotted as an arrow for a galaxy and as an
arrow plus a diamond symbol for a star. The large plotted symbol
represents the radio galaxy. The solid lines display the colour--magnitude}
\end{figure*}

\addtocounter{figure}{-1}
\begin{figure*}
\centerline{
\psfig{file=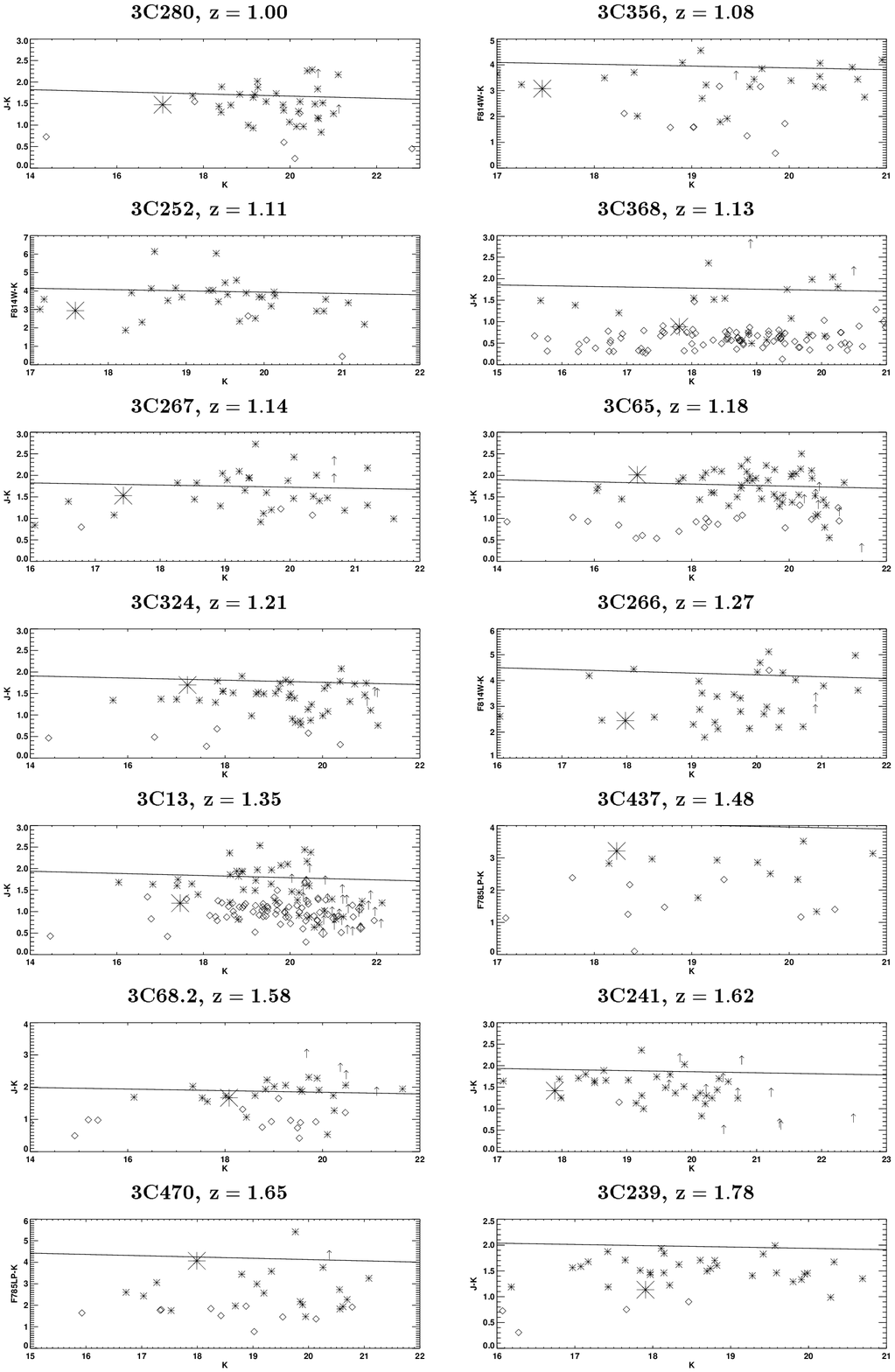,width=14.9cm,clip=}  
}
\caption{{\bf (continued)} relations that would be obtained by simply
passively evolving the colour--magnitude relation observed in the Coma
cluster back in redshift, assuming an elliptical galaxy which formed its
stars in a short time-span at $z=10$.}
\end{figure*}

The colour--magnitude relation for the Coma cluster has been redshifted
and evolved according to the passive evolutionary models for elliptical
galaxies described in the previous section; this redshifted relation is
over-plotted on the figures. Note that the radio galaxies generally have
near--infrared colours similar to this theoretical line, indicating that
these are old elliptical galaxies (e.g. Best \etal\ 1998b)\nocite{bes98d},
but not in all cases: some of the radio galaxies appear redder because a
heavily reddened nuclear component contributes to the K--band emission
(e.g. 3C22, 3C41; Economou \etal\ 1995, Rawlings \etal\ 1995, Best \etal\
1998)\nocite{raw95,eco95,bes98d}, whilst for others the bluer filter is
significantly contaminated by excess optical--UV emission induced by the
radio source (the `alignment effect', e.g. McCarthy \etal\
1987).\nocite{mcc87}

A number of features are apparent from these near--infrared
colour--magnitude diagrams.  Considering initially those radio galaxies
with redshifts $z \lta 0.9$, many of the sample show reasonably convincing
evidence for associated clusters; here it should be born in mind that for
an Abell Class 1 cluster only 10 to 15 associated cluster galaxies are
expected in the observed region of sky down to three magnitudes below the
magnitude of the radio galaxy (see Section~\ref{spatcorr}). Some fields
(e.g. 3C34, 3C337) clearly show at least this number (the `background'
counts at these K--magnitudes and colours are small, as can be seen by a
comparison with the colour--magnitude relations of the higher redshift
sources), whilst other fields (e.g. 3C217) show few if any associated
galaxies.  There are clearly large source--to--source variations in
environmental richness.

Where a clear colour--magnitude relation is observed, the mean colour of
this relation lies close to that calculated theoretically by just
passively evolving the Coma colour--magnitude relations back in redshift,
and the scatter around the colour--magnitude sequence is small. These
results have been shown before for optical and X--ray selected clusters at
redshifts out to $z \sim 0.8$ (e.g. Stanford \etal\ 1998\nocite{sta98}).
There are some small deviations in the colour of the observed sequence
from the passive evolution relation (e.g. see 3C265), but none much larger
than a couple of tenths of a magnitude. The near--infrared colours of the
galaxies in the radio galaxy fields are therefore consistent with those
observed in other clusters at the same redshift, implying that the excess
galaxy counts are associated with a structure at the radio galaxy
redshift. This is important because of previous suggestions that powerful
distant radio galaxies may be systematically amplified by foreground
lensing structures (Le F{\`e}vre \etal\ 1987; Ben{\'\i}tez,
Mart{\'\i}nez--Gonz{\'a}les and Martin--Mirones
1997)\nocite{fev87,ben97a}; were such structures to be present, they could
account for both the excess K--band counts and the peak in the
cross--correlation statistics, but the colour--magnitude relations argue
against this.

At higher redshifts the evidence for clear colour--magnitude relations is
much poorer. This is mainly because the combination of the greatly
increased contribution from field galaxies at these fainter magnitudes and
the increased scatter in the colours due to photometric uncertainties,
results in any colour--magnitude sequence appearing much less
prominent. The difficulty of selecting cluster candidates at these
redshifts on the basis of a single colour can be gauged by examining
3C324; for this field, Dickinson \shortcite{dic97a} has confirmed the
presence of a poor cluster of galaxies, but this is barely apparent from
its colour--magnitude relation. 

Although no prominent colour--magnitude sequences are seen in the higher
redshift fields, there remains a net excess of K--band counts: if the
fields of the radio galaxies are divided into two redshift bins, no
significant differences in the faint galaxy counts are seen between the
high and low redshift fields. A simple analysis also shows that the excess
in the fields of the higher redshift radio galaxies is comprised of red
($J-K \ge 1.75$) galaxies. In Table~\ref{colfracs} are given the mean
number of galaxies per field with magnitudes $17 < K < 20$ and colours
$J-K \ge 1.75$ or $1.25 < J-K < 1.75$, in the low and high redshift
bins. There are more galaxies with bluer $J-K$ colours in the fields of
the lower redshift radio galaxies than in those at higher redshifts, due
to the associated cluster galaxies in the lower redshift fields which have
colours $1.25 \lta J-K \lta 1.75$ (cf. Figure~\ref{redcolfigs}). On the
other hand, there are more galaxies with $J - K \ge 1.75$ colours in the
high redshift than low redshift fields. The excess K--band galaxy counts
in the high redshift field appear to be predominantly associated with red
galaxies, with colours similar to those expected for old cluster galaxies
at these redshifts. This again indicates that the excess number counts are
associated with a structure at the redshift of the radio galaxy rather
than a foreground structure.

\begin{table}
\caption{\label{colfracs} The mean number of galaxies per field with
magnitudes $17 < K < 20$ and blue/red colours, as a function of redshift.}
\begin{tabular}{cccc}
          & No. fields &\multicolumn{2}{c}{~~~Mean number of galaxies} \\
          &            & $1.25 < J-K < 1.75$ & $J-K \ge 1.75$ \\
$z < 0.9$ &   9        &  $19.6 \pm 2.5$     &  $8.1 \pm 1.7$ \\
$z > 0.9$ &   11       &  $14.1 \pm 1.8$     &  $13.1 \pm 2.6$ \\
\end{tabular}
\end{table}

\subsection{Multi--colour relations}

\begin{figure*}
\centerline{
\psfig{file=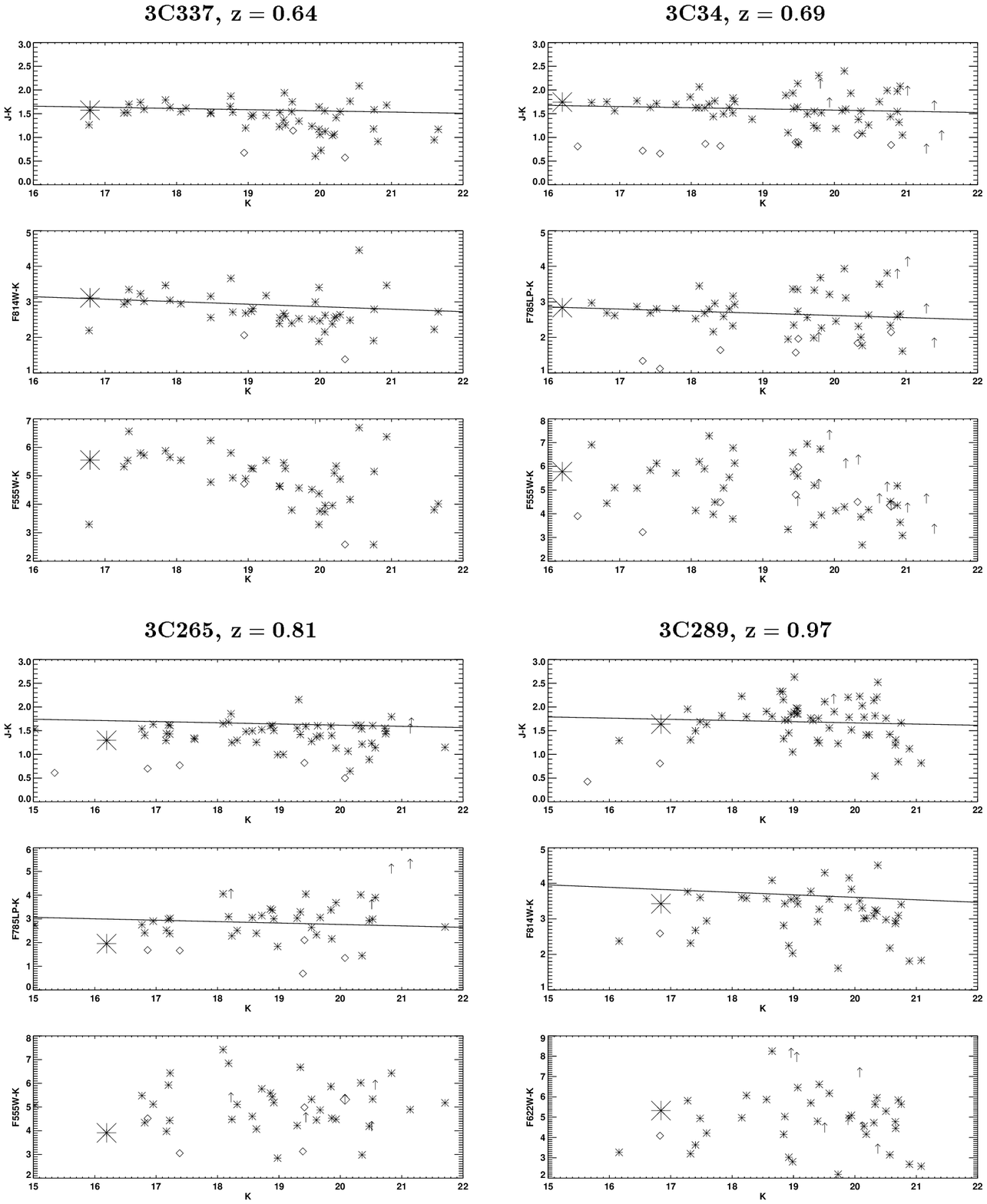,width=\textwidth,clip=}
}
\caption{\label{colmagfigs} Colour--magnitude relations for the 3CR radio
galaxies from the UKIRT and HST data. Symbols as in
Figure~\ref{redcolfigs}. The redshifted Coma colour--magnitude relations
are only plotted on figures for which both filters are longward of the
rest--frame 4000\AA\ break, since shortward of that wavelength small
quantities of recent or on--going star formation can have large effects on
the colours leading to only poorly defined sequences.}
\end{figure*}

\addtocounter{figure}{-1}
\begin{figure*}
\centerline{
\psfig{file=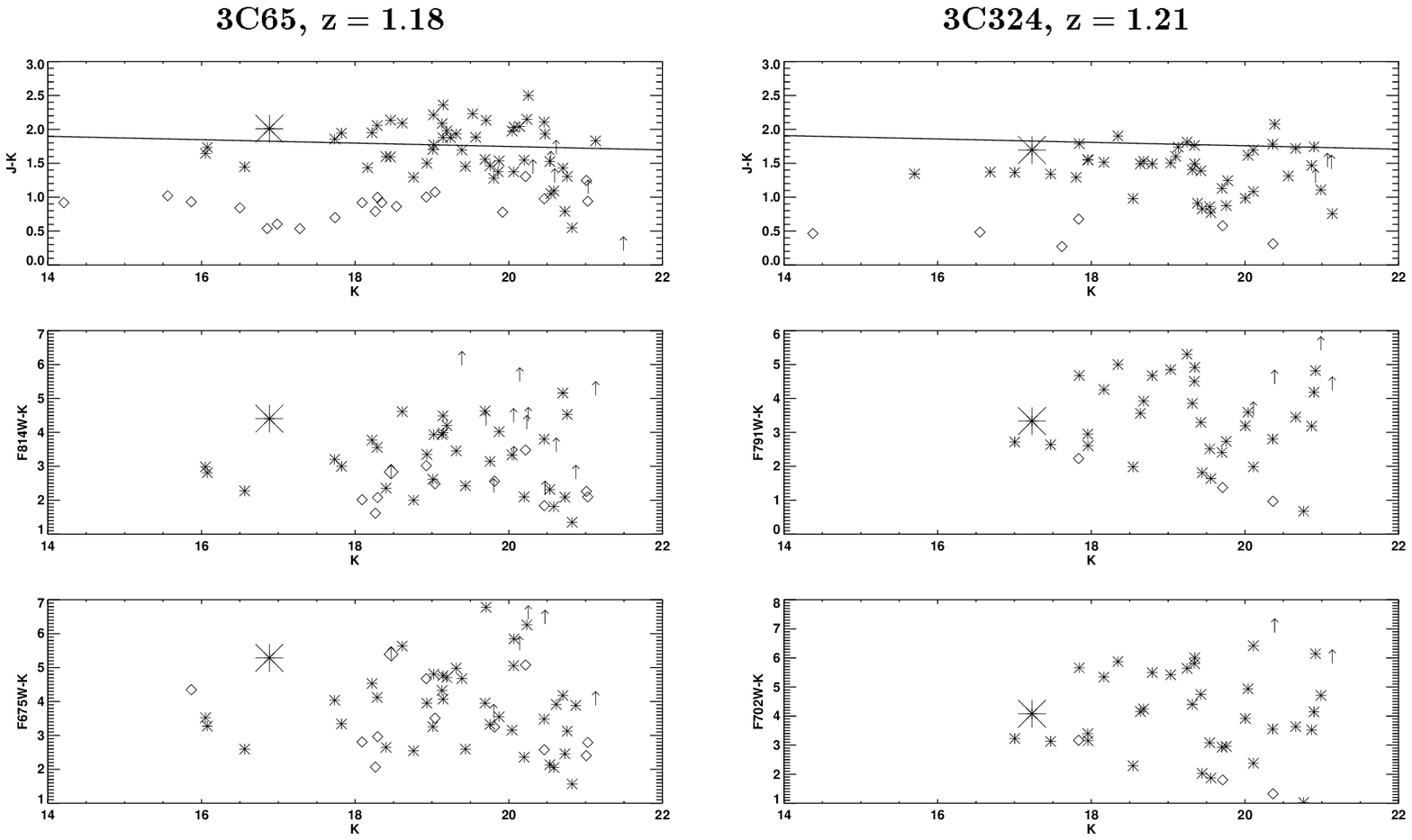,width=\textwidth,clip=}
}
\caption{{\bf cont.}}
\end{figure*}

In Figure~\ref{colmagfigs} are plotted a complete set of colour--magnitude
relations for six galaxies in the sample, chosen to be those which show
amongst the best examples of near--infrared colour--magnitude relations
for their redshift\footnote{Equivalent plots for the other fields are
available from the author on request, but are not reproduced here to save
space.}. From these it is apparent that, although the scatter around the
near--infrared colour--magnitude sequence remains small even at these high
redshifts, colours that reach shortward of the rest--frame 4000\AA\ break
show a dramatic increase in the scatter of the relation.  (e.g. compare
the various relations for 3C34); these colours can be strongly influenced
by small amounts of recent or on--going star formation, indicating that
this may be common in these high redshift clusters. Whether such star
formation is in some way connected with the presence of a powerful radio
source in these clusters cannot be distinguished from these data.

\begin{figure*}
\centerline{
\psfig{file=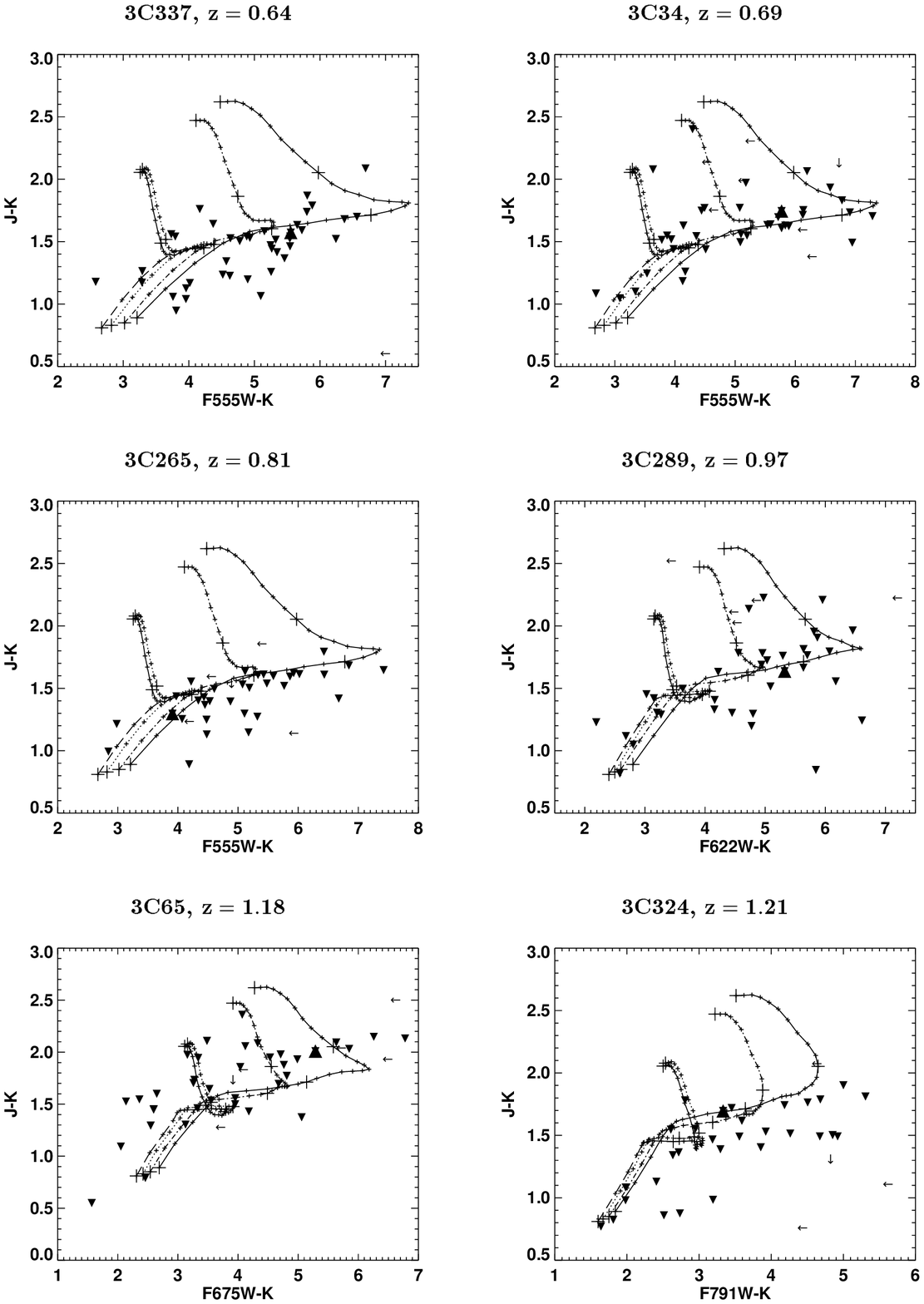,width=15.5cm,clip=}  
}
\caption{\label{colcolplots} Colour--colour plots for fields of the six
radio galaxies presented in Figure~\ref{colmagfigs}. The large triangle
represents the radio galaxy, and the smaller upside-down triangles
represent the other galaxies in the field. Galaxies with upper limits on
their colours are indicated by arrows. The four tracks show the colour
evolution of the passively evolving galaxy models described in
Section~\ref{spatcorr}: E's -- solid line; Sab's -- dot-dash line; Sc's
-- dotted line; Sdm's -- long dashed line. The tracks run from $z=0$
(lower left of plot) to $z=3$, with the small crosses along each line
representing steps of 0.1 in redshift, and redshifts $z=0,1,2,3$ being
indicated by the larger crosses.}
\end{figure*}

To properly investigate the nature of the galaxies in these fields, all of
the colour information must be used simultaneously to derive photometric
redshifts and investigate star formation activity. This is beyond the
scope of the current paper but will be addressed later (Kodama \& Best, in
preparation). Here, the use of the all multi--colour information
simulataneously is merely demonstrated in Figure~\ref{colcolplots} through
colour--colour plots for these six fields. For each field, the
near--infrared $J-K$ colour of each galaxy is plotted against its
optical-infrared colour. These data are then compared against theoretical
evolutionary tracks for the four different passively evolving galaxy
models considered in Section~\ref{spatcorr} (E's, Sab's, Sc's, Sdm's).

For the lower redshift radio galaxies there is clearly a large
concentration of galaxies with colours very close to those of the model
elliptical galaxy at the redshift of the radio source. Further, there is a
distribution of galaxies with colours between this and the colours of the
model spiral galaxies at that redshift. In contrast, for the higher
redshift radio galaxies, no strong concentration of galaxies is seen close
to the elliptical galaxy prediction, and the number of cluster candidates
lying between the locations of the elliptical and spiral model galaxies is
smaller than that found in the lower redshift cases. Clearly, despite the
excess $K$--counts and red $J-K$ galaxies in the fields of these high
redshift radio sources, to accurately investigate cluster membership
requires the construction of photometric redshifts using several colours
measured with high photometric accuracy.

\section{Discussion}
\label{discuss}

In this paper a number of pieces of evidence for clustering around distant
3CR radio galaxies have been presented. These can be summarised as
follows:

\begin{itemize}
\item The K--band number counts show an overdensity of faint galaxies in
the fields of the radio galaxies, with a mean value of 11 excess galaxies
per field.

\item This excess is comparable to the galaxy overdensity expected for a
field of view of this size centred on a cluster of approximately Abell
Class 0 richness.

\item Cross--correlation analyses show a pronounced peak in the angular
cross--correlation function around the radio galaxies. 

\item Assuming that the galaxy luminosity function undergoes pure passive
luminosity evolution with redshift, the corresponding spatial
cross--correlation amplitude lies between those determined for Abell Class
0 and Abell Class 1 clusters.

\item The galaxies in the fields of most of the lower redshift radio
galaxies in the sample show clear near--infrared colour--magnitude
relations with only small scatter. The colours of these sequences are in
agreement with those of other clusters at these redshifts, indicating that
the excess number counts and cross--correlation peak are both associated
with a structure at the redshift of the radio galaxy.

\item There is considerably more scatter in the relations involving
shorter wavelength colours, suggesting low levels of recent or on--going
star formation in many of the galaxies.

\item At higher redshifts the colour--magnitude relations are less
prominent due to increased background contributions, but there is a clear
excess of galaxies with very red infrared colours.
\end{itemize}

These features all provide strong evidence that distant radio galaxies
tend to reside in rich environments.  The number counts, the
cross--correlation statistics, and the colour--magnitude relations all
complement the previous results from X--ray imaging, narrow--band imaging,
spectroscopic studies and radio polarisation studies discussed in
Section~\ref{intro}. A coherent picture that most, but not all, high
redshift radio sources live in poor to medium richness clusters has now
been built.

Taking the results at face value, the environmental richness around these
$z \sim 1$ radio galaxies is higher than that around powerful radio
galaxies at $z \sim 0.5$ calculated by Yates \etal\ \shortcite{yat89} and
by Hill and Lilly \shortcite{hil91}.  This suggests that the increase
between $z=0$ and $z=0.5$ in the mean richness of the environments
surrounding FR\,II radio galaxies, found by those authors, continues to
higher redshifts.

Some notes of caution must be added to this conclusion. First, from the
variations in richness of the colour--magnitude relation at any given
redshift, (e.g. compare 3C217 and 3C226), it is apparent that there is a
wide spread in the density of the environments in these fields. Although
most show some evidence of living in at least group environments, not all
powerful distant radio galaxies lie in clusters. Second, a simple visual
comparison with the extremely rich high redshift cluster MS1054$-$03
($z=0.83$, van Dokkum 1999)\nocite{dok99b} is sufficient to demonstrate
that even at high redshifts, powerful radio galaxies still avoid the most
extreme richness clusters.

Further, although the galaxy count excesses and the cross--correlation
amplitudes have been compared with those of Abell clusters at low
redshifts, in hierarchical galaxy formation models such comparisons will
always be somewhat ambiguous. On--going mergers between galaxies mean that
more sub--clumps are seen at higher redshifts, whilst the general galaxy
cross--correlation length also evolves with redshift in a manner dependent
upon both cosmological parameters and the method of selecting the galaxy
populations \cite{kau99}. Therefore, quantitative interpretations of
either of these parameters at high redshift must be considered with some
care. On the other hand, in hierarchical growth models, the structures in
which the radio galaxies reside will also continue to grow and evolve into
much richer structures by a redshift of zero, meaning that the qualitative
result that the high redshift radio galaxies lie in rich environments
for their redshift is secure.
\smallskip

Finally, it is important to consider the consequences for our
understanding of the onset and nature of powerful radio sources of a
change in the environments of FR\,II radio galaxies from groups at low
redshifts to clusters at high redshift. As discussed by Best \etal\
\shortcite{bes98d}, if this result holds then the standard interpretation
of the tightness and slope of the Hubble $K-z$ relation, that of
`closed--box passive evolution' of radio galaxies at $z \gta 1$ into radio
galaxies at $z \sim 0$, is no longer valid. It is not possible that the
environments can become less rich with progressing cosmic time. Instead,
Best \etal\ propose that powerful radio galaxies selected at high and low
redshift have different evolutionary histories but must contain a similar
mass of stars, a few times $10^{11} M_{\odot}$, so conspiring to produce
the observed `passively evolving' K$-z$ relation. In their model, powerful
FR\,II radio sources are seldom formed in more massive galaxies (that is,
in central cluster galaxies at low redshifts) because of the difficulties
in supplying sufficient fueling gas to the black hole: in rich low
redshift clusters the galaxies and gas have been virialised and take up
equilibrium distributions within the cluster gravitational potential, the
galaxies have high velocity dispersions greatly reducing the merger
efficiency, and there is a dearth of gas--rich galaxies close to the
centre of the clusters which might merge with, and fuel, the central
galaxy. Thus, the formation of a powerful radio source in these
environments is a rare event (but can still happen, e.g. Cygnus A).

At high redshifts, radio galaxies can be found in (proto) cluster
environments because these are not yet virialised, have frequent galaxy
mergers, and have a plentiful supply of disturbed intracluster gas to fuel
the central engine and confine the radio lobes. The central cluster
galaxies will be amongst the most massive galaxies at these redshifts and
so, from the correlation between black hole mass and bulge mass
(e.g. Kormendy and Richstone 1995)\nocite{kor95}, will have the most
massive black holes. The kinetic energy of the relativistic radio jets of
distant 3CR radio galaxies corresponds to the Eddington luminosity of a
black hole with $M \sim 10^8 -10^9\,M_{\odot}$ \cite{raw91b}, implying
that these sources are fueled close to the Eddington limit. Therefore the
most powerful radio sources will tend to be powered by the most massive
central engines and hence be hosted by the most massive galaxies, which
tend to be found at the centres of forming rich clusters. The significant
scatter in the black hole versus bulge mass correlation \cite{kor95}
would, however, result in some scatter in the richness of the environments
of the radio galaxies. The data presented in this paper are in full
agreement with this model.

\section*{Acknowledgements} 

This work was supported in part by the Formation and Evolution of Galaxies
network set up by the European Commission under contract ERB FMRX--
CT96--086 of its TMR programme. The United Kingdom Infrared Telescope is
operated by the Joint Astronomy Centre on behalf of the U.K. Particle
Physics and Astronomy Research Council. This work is, in part, based on
observations made with the NASA/ESA Hubble Space Telescope, obtained at
the Space Telescope Science Institute, which is operated by AURA Inc.,
under contract from NASA. I thank Huub R{\"o}ttgering for useful
discussions, and the referee for helpful comments.

\label{lastpage}
\bibliography{pnb} 

\begin{thebibliography}{}

\bibitem[\protect\citename{Andersen \& Owen{\ }}{1994}]{and94}
Andersen~V.,  Owen~F.~N.,  1994, AJ, 108, 361

\bibitem[\protect\citename{Barthel \& Arnaud{\ }}{1996}]{bar96a}
Barthel~P.~D.,  Arnaud~K.~A.,  1996, MNRAS, 283, L45

\bibitem[\protect\citename{Ben{\'\i}tez et~al.{\ }}{1997}]{ben97a}
Ben{\'\i}tez~N.,  Mart{\'\i}nez-Gonz{\'a}les~E.,    Martin-Mirones~J.~M.,
  1997, A\&A, 321, L1

\bibitem[\protect\citename{Bershady et~al.{\ }}{1998}]{ber98}
Bershady~M.~A.,  Lowenthal~J.~D.,    Koo~D.~C.,  1998, ApJ, 505, 50

\bibitem[\protect\citename{Bertin \& Arnouts{\ }}{1996}]{ber96}
Bertin~E.,  Arnouts~S.,  1996, A\&A Supp., 117, 393

\bibitem[\protect\citename{Best et~al.{\ }}{1998a}]{bes98a}
Best~P.~N.,  Carilli~C.~L.,  Garrington~S.~T.,  Longair~M.~S.,
  R{\"o}ttgering~H. J.~A.,  1998a, MNRAS, 299, 357

\bibitem[\protect\citename{Best et~al.{\ }}{1997}]{bes97c}
Best~P.~N.,  Longair~M.~S.,    R{\"o}ttgering~H. J.~A.,  1997, MNRAS, 292, 758

\bibitem[\protect\citename{Best et~al.{\ }}{1998b}]{bes98d}
Best~P.~N.,  Longair~M.~S.,    R{\"o}ttgering~H. J.~A.,  1998b, MNRAS, 295, 549

\bibitem[\protect\citename{Bower et~al.{\ }}{1998}]{bow98}
Bower~R.~G.,  Kodama~T.,    Terlevich~A.,  1998, MNRAS, 299, 1193

\bibitem[\protect\citename{Bruzual \& Charlot{\ }}{1993}]{bru93}
Bruzual~G.,  Charlot~S.,  1993, ApJ, 405, 538

\bibitem[\protect\citename{Burstein \& Heiles{\ }}{1982}]{bur82a}
Burstein~D.,  Heiles~C.,  1982, AJ, 87, 1165

\bibitem[\protect\citename{Butcher \& Oemler{\ }}{1978}]{but78}
Butcher~H.~R.,  Oemler~A.,  1978, ApJ, 219, 18

\bibitem[\protect\citename{Carilli et~al.{\ }}{1997}]{car97}
Carilli~C.~L.,  R{\"o}ttgering~H. J.~A.,  {van Ojik}~R.,  Miley~G.~K.,    {van
  Breugel}~W. J.~M.,  1997, ApJ Supp., 109, 1

\bibitem[\protect\citename{Crawford \& Fabian{\ }}{1996}]{cra96b}
Crawford~C.~S.,  Fabian~A.~C.,  1996, MNRAS, 282, 1483

\bibitem[\protect\citename{Dickinson{\ }}{1997}]{dic97a}
Dickinson~M.,  1997, in Tanvir~N.~R.,  Arag{\'o}n-Salamanca~A.,   Wall~J.~V.,
  eds, HST and the high redshift Universe.
Singapore: World Scientific, p.~207

\bibitem[\protect\citename{Djorgovski \& Davies{\ }}{1987}]{djo87b}
Djorgovski~S.,  Davies~M.,  1987, ApJ, 313, 59

\bibitem[\protect\citename{Djorgovski et~al.{\ }}{1995}]{djo95}
Djorgovski~S.,  Soifer~B.~T.,  Pahre~M.~A.,  Larkin~J.~E.,  Smith~J.~D.,
  Neugebauer~G.,  Smail~I.,  Matthews~K.,  Hogg~D.~W.,  Blandford~R.~D.,
  Cohen~J.,  Harrison~W.,    Nelson~J.,  1995, ApJ, 438, L13

\bibitem[\protect\citename{Dressler et~al.{\ }}{1987}]{dre87a}
Dressler~A.,  {Lyndon-Bell}~D.,  Burstein~D.,  Davies~R.~L.,  Faber~S.~M.,
  Terlevich~R.,    Wegner~G.,  1987, ApJ, 313, 42

\bibitem[\protect\citename{Dunlop et~al.{\ }}{1996}]{dun96}
Dunlop~J.~S.,  Peacock~J.,  Spinrad~H.,  Dey~A.,  Jimenez~R.,  Stern~D.,
  Windhorst~R.,  1996, Nat, 381, 581

\bibitem[\protect\citename{Economou et~al.{\ }}{1995}]{eco95}
Economou~F.,  Lawrence~A.,  Ward~M.~J.,    Blanco~P.~R.,  1995, MNRAS, 272, L5

\bibitem[\protect\citename{Fanaroff \& Riley{\ }}{1974}]{fan74}
Fanaroff~B.~L.,  Riley~J.~M.,  1974, MNRAS, 167, 31P

\bibitem[\protect\citename{Gardner et~al.{\ }}{1993}]{gar93}
Gardner~J.~P.,  Cowie~L.~L.,    Wainscoat~R.~J.,  1993, ApJ, 415, L9

\bibitem[\protect\citename{Giavalisco et~al.{\ }}{1998}]{gia98}
Giavalisco~M.,  Steidel~C.~C.,  Adelberger~K.~L.,  Dickinson~M.~E.,
  Pettini~M.,    Kellogg~M.,  1998, ApJ, 503, 543

\bibitem[\protect\citename{Guiderdoni \& {Rocca--Volmerange}{\ }}{1990}]{gui90}
Guiderdoni~B.,  {Rocca--Volmerange}~B.,  1990, A\&A, 227, 362

\bibitem[\protect\citename{Hill \& Lilly{\ }}{1991}]{hil91}
Hill~G.~J.,  Lilly~S.~J.,  1991, ApJ, 367, 1

\bibitem[\protect\citename{Holtzman et~al.{\ }}{1995}]{hol95}
Holtzman~J.~A.,  Burrows~C.~J.,  Casertano~S.,  Hester~J.~J.,  Trauger~J.~T.,
  Watson~A.~M.,    Worthey~G.,  1995, PASP, 107, 1065

\bibitem[\protect\citename{Kauffmann \& Charlot{\ }}{1998}]{kau98a}
Kauffmann~G.,  Charlot~S.,  1998, MNRAS, 297, L23

\bibitem[\protect\citename{Kauffmann et~al.{\ }}{1999}]{kau99}
Kauffmann~G.,  Colberg~J.~M.,  Diaferio~A.,    White~S. D.~M.,  1999, MNRAS,
  307, 529

\bibitem[\protect\citename{Kormendy \& Richstone{\ }}{1995}]{kor95}
Kormendy~J.,  Richstone~D.,  1995, ARA\&A, 33, 581

\bibitem[\protect\citename{Kron{\ }}{1980}]{kro80}
Kron~R.~G.,  1980, ApJ Supp., 43, 305

\bibitem[\protect\citename{Kurk et~al.{\ }}{2000}]{kur00b}
Kurk~J.~D.,  R{\"o}ttgering~H. J.~A.,  Pentericci~L.,    Miley~G.~K.,  2000,
  A\&A: submitted.

\bibitem[\protect\citename{Laing et~al.{\ }}{1983}]{lai83}
Laing~R.~A.,  Riley~J.~M.,    Longair~M.~S.,  1983, MNRAS, 204, 151

\bibitem[\protect\citename{Lambas et~al.{\ }}{1992}]{lam92}
Lambas~D.~G.,  Maddox~S.~J.,    Loveday~J.,  1992, MNRAS, 258, 404

\bibitem[\protect\citename{Lauer{\ }}{1989}]{lau89}
Lauer~T.~R.,  1989, PASP, 101, 445

\bibitem[\protect\citename{{Le F{\`e}vre} et~al.{\ }}{1987}]{fev87}
{Le F{\`e}vre}~O.,  Hammer~F.,  Nottale~L.,    Mathez~G.,  1987, Nat, 326, 268

\bibitem[\protect\citename{Longair \& Seldner{\ }}{1979}]{lon79b}
Longair~M.~S.,  Seldner~M.,  1979, MNRAS, 189, 433

\bibitem[\protect\citename{Loveday{\ }}{2000}]{lov00}
Loveday~J.,  2000, MNRAS: in press.

\bibitem[\protect\citename{Mao et~al.{\ }}{1998}]{mao98}
Mao~S.,  Mo~H.~J.,    White~S.~D.,  1998, MNRAS, 297, L71

\bibitem[\protect\citename{McCarthy et~al.{\ }}{1995}]{mcc95}
McCarthy~P.~J.,  Spinrad~H.,    {van Breugel}~W. J.~M.,  1995, ApJ Supp., 99,
  27

\bibitem[\protect\citename{McCarthy et~al.{\ }}{1987}]{mcc87}
McCarthy~P.~J.,  {van Breugel}~W. J.~M.,  Spinrad~H.,    Djorgovski~S.,  1987,
  ApJ, 321, L29

\bibitem[\protect\citename{Mcleod et~al.{\ }}{1995}]{mcl95}
Mcleod~B.~A.,  Bernstein~G.~M.,  Rieke~M.~J.,  Tollestrup~E.~V.,
  Fazio~G.~G.,  1995, ApJ Supp., 96, 117

\bibitem[\protect\citename{Minezaki et~al.{\ }}{198}]{min98}
Minezaki~T.,  Kobayashi~Y.,  Yoshii~Y.,    Peterson~B.~A.,  198, ApJ, 494, 111

\bibitem[\protect\citename{Mobasher et~al.{\ }}{1993}]{mob93}
Mobasher~B.,  Sharples~R.~M.,    Ellis~R.~S.,  1993, MNRAS, 263, 560

\bibitem[\protect\citename{Moustakas et~al.{\ }}{1997}]{mou97}
Moustakas~L.~A.,  Davis~M.,  Graham~J.~R.,  Silk~J.,  Peterson~B.~A.,
  Yoshii~Y. L.~M.,  Coles~P.,  Lucchin~F.,    Matarrese~S.,  1997, ApJ, 475,
  445

\bibitem[\protect\citename{Owen et~al.{\ }}{1997}]{owe97}
Owen~F.~N.,  Ledlow~M.~J.,  Morrison~G.~E.,    Hill~J.~M.,  1997, ApJ, 488, L15

\bibitem[\protect\citename{Prestage \& Peacock{\ }}{1988}]{pre88}
Prestage~R.~M.,  Peacock~J.~A.,  1988, MNRAS, 230, 131

\bibitem[\protect\citename{Rawlings et~al.{\ }}{1995}]{raw95}
Rawlings~S.,  Lacy~M.,  Sivia~D.~S.,    Eales~S.~A.,  1995, MNRAS, 274, 428

\bibitem[\protect\citename{Rawlings \& Saunders{\ }}{1991}]{raw91b}
Rawlings~S.,  Saunders~R.,  1991, Nat, 349, 138

\bibitem[\protect\citename{Roche et~al.{\ }}{1998}]{roc98a}
Roche~N.,  Eales~S.,    Hippelein~H.,  1998, MNRAS, 295, 946

\bibitem[\protect\citename{Rosati et~al.{\ }}{1998}]{ros98}
Rosati~P.,  {Della Ceca}~R.,  Norman~C.,    Giacconi~R.,  1998, ApJ, 492, L21

\bibitem[\protect\citename{Schechter{\ }}{1976}]{sch76}
Schechter~P.~L.,  1976, ApJ, 203, 297

\bibitem[\protect\citename{Scodeggio et~al.{\ }}{1999}]{sco99}
Scodeggio~M.,  Olsen~L.~F.,  {da Costa}~L.,  Slijkhuis~R.,  Deul~E.,  Erben~T.,
   Hook~R.,  Nonino~M.,  Wicenec~A.,    Zaggia~S.,  1999, A\&A Supp., 137, 83

\bibitem[\protect\citename{Simard et~al.{\ }}{1999}]{sim99c}
Simard~L.,  Koo~D.~C.,  Faber~S.~M.,  Sarajedini~V.~L.,  Vogt~N.~P.,
  Phillips~A.~C.,  Gebhardt~K.,  Illingworth~G.~D.,    Wu~K.~L.,  1999, ApJ,
  519, 563

\bibitem[\protect\citename{Songaila et~al.{\ }}{1994}]{son94}
Songaila~A.,  Cowie~L.~L.,  Hu~E.~M.,    Gardner~J.~P.,  1994, ApJ Supp., 94,
  461

\bibitem[\protect\citename{Stanford et~al.{\ }}{1998}]{sta98}
Stanford~S.~A.,  Eisenhardt~P.~R.,    Dickinson~M.,  1998, ApJ, 492, 461

\bibitem[\protect\citename{Stanford et~al.{\ }}{1997}]{sta97}
Stanford~S.~A.,  Elston~R.,  Eisenhardt~P.~R.,  Spinrad~H.,  Stern~D.,
  Dey~A.,  1997, AJ, 114, 2232

\bibitem[\protect\citename{Szokoly et~al.{\ }}{1998}]{szo98}
Szokoly~G.~P.,  Subbarao~M.~U.,  Connolly~A.~J.,    Mobasher~B.,  1998, ApJ,
  492, 452

\bibitem[\protect\citename{{van Breugel} et~al.{\ }}{1999}]{bre99}
{van Breugel}~W. J.~M.,  {De Breuck}~C.,  Stanford~S.~A.,  Stern~D.,
  R{\"o}ttgering~H. J.~A.,    Miley~G.~K.,  1999, ApJ, 518, L61

\bibitem[\protect\citename{{van Dokkum}{\ }}{1999}]{dok99b}
{van Dokkum}~P.~G.,  1999, Ph.D. thesis, University of Groningen

\bibitem[\protect\citename{{van Dokkum} et~al.{\ }}{1999}]{dok99a}
{van Dokkum}~P.~G.,  Franx~M.,  Fabricant~D.,  Kelson~D.~D.,
  Illingworth~G.~D.,  1999, ApJ, 520, L95

\bibitem[\protect\citename{White \& Frenk{\ }}{1991}]{whi91}
White~S. D.~M.,  Frenk~C.~S.,  1991, ApJ, 379, 52

\bibitem[\protect\citename{Wold et~al.{\ }}{2000}]{wol00}
Wold~M.,  Lacy~M.,  Lilje~P.~B.,    Serjeant~S.,  2000, MNRAS: submitted.,
  astro-ph/9912070

\bibitem[\protect\citename{Yates et~al.{\ }}{1989}]{yat89}
Yates~M.~G.,  Miller~L.,    Peacock~J.~A.,  1989, MNRAS, 240, 129

\bibitem[\protect\citename{Yee \& {L{\'o}pez--Cruz}{\ }}{1999}]{yee99}
Yee~H. K.~C.,  {L{\'o}pez--Cruz}~O.,  1999, AJ, 117, 1985

\end{thebibliography}
\bibliographystyle{mn} 

\end{document}